\documentclass[12pt]{article}
\usepackage[a4paper,margin=1in]{geometry}
\usepackage[titletoc,title]{appendix}
\usepackage{changepage}
\usepackage[utf8x]{inputenc}
\usepackage{textcomp,marvosym}
\usepackage{fixltx2e}
\usepackage{amsmath,amssymb}
\usepackage{cite}
\usepackage{nameref,hyperref,nccmath}
\usepackage{authblk}
\usepackage{algorithm}
\usepackage{algpseudocode}
\usepackage{dsfont}
\usepackage{xcolor}
\usepackage{enumitem}
\usepackage{caption,setspace}
\usepackage{subcaption}
\usepackage{pdflscape}
\usepackage{graphicx}
\usepackage{placeins}
\usepackage{multirow,mathtools,nccmath}
\DeclareMathAlphabet\mathbfcal{OMS}{cmsy}{b}{n}

\usepackage{mathtools}

\renewcommand{\eqref}[1]{Equation~(\ref{#1})}

\renewcommand{\baselinestretch}{1.5}

\title{\sc{Fisher-KPP-type models of biological invasion: Open source computational tools,  key concepts and analysis}}

\author[1]{Matthew~J. Simpson\footnote{To whom correspondence should be addressed. E-mail: matthew.simpson@qut.edu.au}}
\author[1]{Scott~W. McCue}

\affil[1]{School of Mathematical Sciences, Queensland University of Technology, Brisbane, Queensland 4001, Australia.}

\begin{document}

\maketitle
\begin{abstract}
This review provides open-access computational tools that support a range of mathematical approaches to analyse three related scalar reaction-diffusion models used to study biological invasion.  Starting with the classic Fisher-Kolmogorov (Fisher-KPP) model, we illustrate how computational methods can be used to explore time-dependent partial differential equation (PDE) solutions in parallel with phase plane and regular perturbation techniques to explore invading travelling wave solutions moving with dimensionless speed $c \ge 2$.  To overcome the lack of a well-defined sharp front in solutions of the Fisher-KPP model, we also review two alternative modeling approaches.  The first is the Porous-Fisher model where the linear diffusion term is replaced with a degenerate nonlinear diffusion term.  Using phase plane and regular perturbation methods, we explore the distinction between sharp- and smooth-fronted invading travelling waves that move with dimensionless speed $c \ge 1/\sqrt{2}$. The second alternative approach is to reformulate the Fisher-KPP model as a moving boundary problem on $0 < x < L(t)$, leading to the Fisher-Stefan model with sharp-fronted travelling wave solutions arising from a PDE model with a linear diffusion term.  Time-dependent PDE solutions and phase plane methods show that travelling wave solutions of the Fisher-Stefan model can describe both biological invasion $(c > 0)$ and biological recession $(c < 0)$.  Open source Julia code to replicate all computational results in this review is available on \href{https://github.com/ProfMJSimpson/PDEInvasion}{GitHub}; we encourage researchers to use this code directly or to adapt the code as required for more complicated models.
\end{abstract}

\newpage
\section{Introduction}\label{sec:Introduction}
\begin{figure}[htp]
    \begin{center}
     \includegraphics[width=1.0\textwidth]{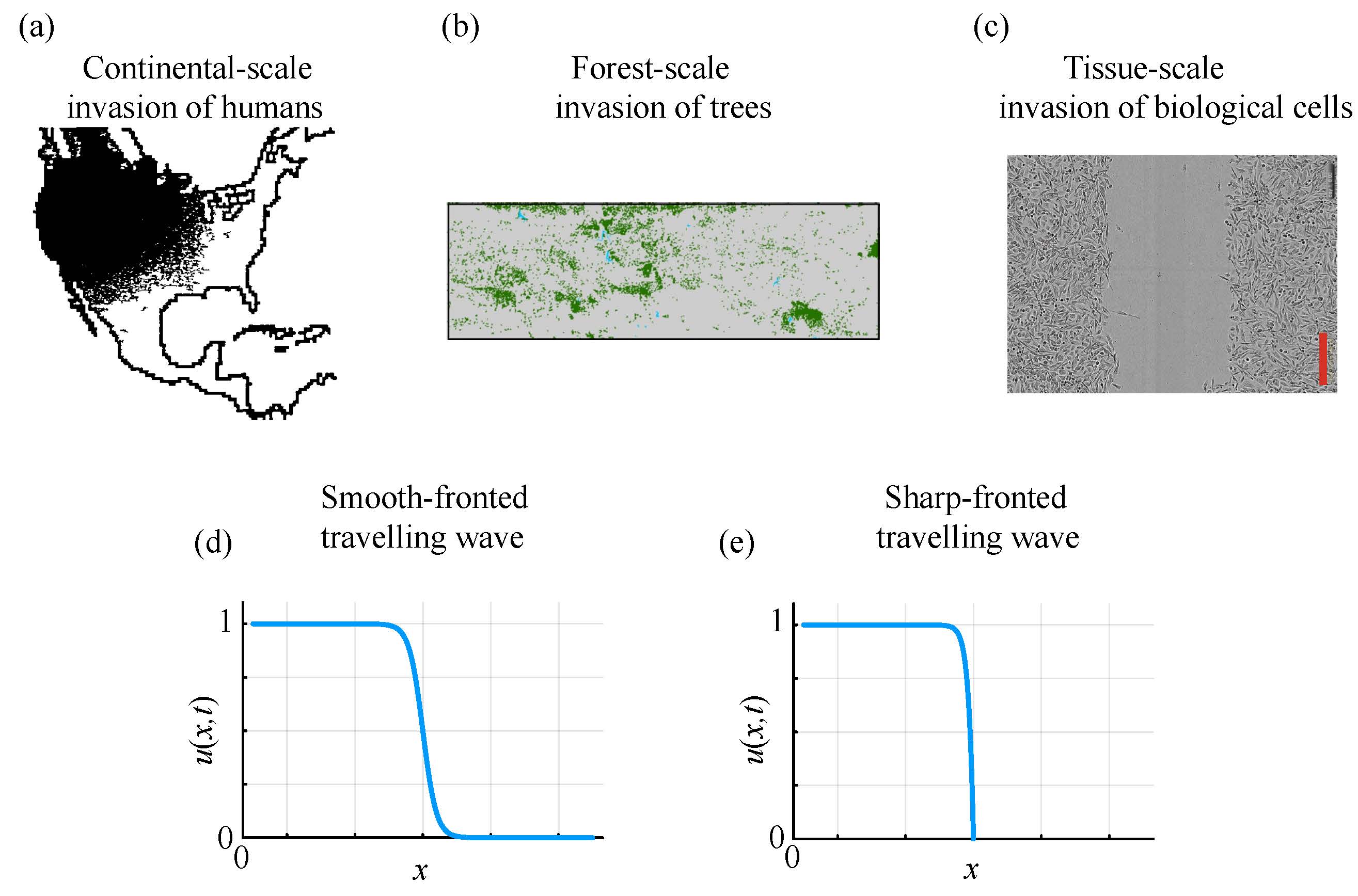}
    \renewcommand{\baselinestretch}{1.0}
     \caption{Biological motivation.  (a) Continental-scale invasion of populations of humans~\cite{Steel1998}. (c) Forest-scale invasion of plants and trees during forest recovery in Puerto Rico~\cite{Acevedo2012}. (d) Tissue-scale growth and expansion of population of malignant cancer cells~\cite{Jin2016,Simpson2024b}.  All three examples have been modelled using simple extensions of the classic Fisher-KPP model. (d) Schematic travelling wave with a smooth front where $u(x,t) > 0$ on $0 < x < \infty$.  (e) Schematic travelling wave with a sharp, well--defined front where $u(x,t) > 0$ for $x < L(t)$, with $x(L(t),t)=0$.} \label{F1}
    \end{center}
\end{figure}

Population biology can involve examining populations of individuals undergoing some form of migration mechanism together with a birth-death process.  The combination of these two individual-level mechanisms can lead to the spatial expansion of that population in the form of moving density fronts.  In certain contexts, these moving fronts are referred to as \textit{biological invasion} or waves of \textit{biological colonisation}.  There are many biological and ecological applications in which biological invasion is important.  These range from continental-scale ($25 \times 10^6$ km$^2$) colonisation of human populations over thousands of years~\cite{Steel1998} (Figure \ref{F1}(a)); forest-scale ($10^5$ hectare) colonisation and growth of tropical rainforests over several decades~\cite{Acevedo2012} (Figure \ref{F1}(b)); and tissue-scale (1-2~mm$^2$) colonisation of populations of biological cells over two days~\cite{Jin2016,Simpson2024b} (Figure \ref{F1}(c)).  As these three very disparate motivating examples indicate, the study of biological invasion, and hence the motivation to use mathematical models to explore and interrogate biological invasion, is particularly important in the field of ecology~\cite{Skellam1951,Shigesada1995,Kot2003} and cell biology~\cite{Murray2002,Edelstein2005}.  In the field of cell biology, in particular, invasion fronts of motile and proliferative cells is important in tissue repair~\cite{Sherratt1990,Maini2004a,Maini2004b,Cai2007,Sengers2007}, embryonic development~\cite{Landman2003,Simpson2006,Simpson2007,DiTalia2022} and disease progression, such as cancer~\cite{Gatenby1996,Perumpanani1999,Swanson2003,Browning2019}.

\newpage
Continuum models of biological invasion are often extensions of the well-known \textit{Fisher-Kolmogorov} (Fisher-KPP) model that was independently proposed by Fisher~\cite{Fisher1937} as well as Kolmogorov, Petrovskii and Piskunov~\cite{Kolmogorov1937} in 1937,
\begin{equation}
\dfrac{\partial \bar{u}}{\partial \bar{t}} = \bar{D} \dfrac{\partial^2 \bar{u}}{\partial \bar{x}^2} + \bar{\lambda} \bar{u}\left[1-\dfrac{\bar{u}}{\bar{K}}\right], \label{Eq:DimensionalFKPP}
\end{equation}
where $\bar{u} \ge 0$ is a dimensional density $[\textrm{individuals}/L^2]$, $\bar{D}>0$ is the diffusivity coefficient $[L^2/T]$ that is proportional to the rate at which individuals in the population undergo random, undirected migration, $\bar{\lambda}>0$ $[/T]$ is a growth rate associated with the logistic source term, and $\bar{K}$ is a carrying capacity density $[\textrm{individuals}/L^2]$.  This simple reaction--diffusion model supposes that random movement of individuals in the population corresponds to a macroscopic linear diffusion term, and that the proliferation of individuals is described by a logistic growth term.  While this model was first proposed as a theoretical modelling tool in 1937, it was not until 1990 that generalisations of this model were first used to explicitly study wound healing processes where $\bar{u}$ represents a density of epithelial cells that close a wound as a result of random migration and carrying capacity-limite logistic proliferation~\cite{Sherratt1990}. While the Fisher-KPP model involves a logistic source term, it is possible to work with a different sigmoid growth function, such as the Richards', Gompertz or generalised logistic models~\cite{Richards1959,Tsoularis2002,Simpson2022}.

An important feature of the Fisher-KPP model, and generalisations thereof, is the ability of these time-dependent partial differential equation (PDE) models to support travelling wave solutions.  Travelling wave solutions are the long-time limit of a time-dependent solution, $\bar{t} \to \infty$~\cite{Canosa1973}, where the time-dependent problem is posed on an infinite (or semi-infinite) domain with appropriate boundary conditions.  The analysis of travelling wave solutions is a key topic within the applied mathematics  literature~\cite{Fife1977,Ablowitz1979,Aronson1980,Bramson1986,vanSaarloos2003,Berestycki2018a,Berestycki2018b,McCue2021}.  Within this context, the  Fisher-KPP model is often viewed as a prototype model of high interest, especially in terms of senior undergraduate and graduate-level training in mathematical sciences~\cite{Kot2003,Murray2002,Edelstein2005}.  While many methods of analysis have been deployed to interrogate and understand travelling wave solutions of the Fisher-KPP model, computational methods for solving the underlying time-dependent PDE model remain of central importance because travelling wave solutions arise as a long-time limit of the time-dependent solution of the nonlinear PDE model.

One well-known feature of travelling wave solutions of the Fisher-KPP model is that the travelling wave solutions are smooth, and do not have compact support as illustrated by the schematic travelling wave profile in Figure \ref{F1}(d).  This feature is at odds with many biological observations that involve a clear front position~\cite{Simpson2013,Johnston2015}.  This limitation of the Fisher-KPP model has been dealt with in two different ways.  The first approach for introducing a sharp-fronted travelling wave solution is to generalise the linear diffusion term to a degenerate porous-media type nonlinear diffusivity~\cite{Pattle1959,Harris04,Vazquez2006,McCue2019}, giving rise to the \textit{Porous-Fisher}  model~\cite{Murray2002,Sengers2007,Pattle1959,Harris04,Vazquez2006,McCue2019,Sanchez1994,Witelski1995,Witelski1997,Sherratt1996,Buenzli2020,Johnston2023,Simpson2024c}. A second way of introducing a sharp-fronted travelling wave solution is to re-formulate the Fisher-KPP model, with linear diffusion and logistic growth, as a moving boundary model on $0 < \bar{x} < \bar{L}(\bar{t})$ together with a Stefan-type boundary condition at $\bar{x} = \bar{L}(\bar{t})$~\cite{Crank1987,Gupta2017}.  This second type of generalisation has been referred to as the \textit{Fisher-Stefan} model of biological invasion~\cite{Du2010,Du2011,ElHachem2019,Elhachem2021,Simpson2020,Elhachem2022,Bui2024}.  As we will show, these three relatively simple scalar reaction-diffusion models give rise to a range of interesting and partly tractable travelling wave solutions that include smooth and sharp-fronted travelling waves as illustrated in Figure \ref{F1}(d)-(e).    Furthermore, these models can be used to draw a distinction between \textit{invading} travelling waves where $u(X,t)$ is an increasing function of time at a fixed location $X$, as well as \textit{receding} travelling waves where $u(X,t)$ is a decreasing function of time at a fixed location $X$.

The aims of this review are broadly two-fold. First, we aim to provide an introduction to the well-known Fisher-KPP (Section \ref{sec:FisherSmooth}) model together with details of the two lesser-known alternatives mentioned above, namely the Porous-Fisher (Section \ref{sec:PorousFisherSmooth}) and Fisher-Stefan (Section \ref{sec:FisherStefan}) models.  As mentioned already, a recurring theme in our discussion will be ability of each model to describe sharp or smooth moving fronts.  While all of these issues have been addressed in the literature, a feature of this review is that we will compare and contrast these three models in a single document that is accessible to graduate students and more experienced researchers alike.  Second we present a series of intuitive and easy-to-use open access computational tools to generate time-dependent PDE solutions for the Fisher-KPP, Porous-Fisher and Fisher-Stefan models.  Using computational tools we relate long-time PDE solutions with associated phase plane analysis and provide open access computational tools to visualise and explore the associated phase planes for travelling wave solutions of each model.  We review and derive a range of various perturbation solutions that provide insight into important features of the travelling wave solutions in various asymptotic limits. Finally, we also provide, in the Appendix, another set of results for a fourth model that poses the Porous-Fisher model as a moving boundary problem which we call the Porous-Fisher-Stefan model~\cite{Fadai2020,Simpson2023,Simpson2024a}.  Again, by providing all of our code written in the open source Julia programming language on \href{https://github.com/ProfMJSimpson/PDEInvasion}{GitHub}, we aim to provide all researchers a suite of clear and practical tools to study PDE models for biological invasion that may be used directly or adapted as required.

\section{Fisher-KPP model: Smooth initial conditions}\label{sec:FisherSmooth}
Starting with the dimensional Fisher-KPP model, Equation (\ref{Eq:DimensionalFKPP}), we re--scale density $u = \bar{u} / \bar{K}$, time $t = \bar{\lambda}\bar{t}$ and space $x = \bar{x} \sqrt{\bar{\lambda}/\bar{D}}$ to give a non-dimensional PDE model,
\begin{equation}
\dfrac{\partial u}{\partial t} = \dfrac{\partial^2 u}{\partial x^2} + u(1-u), \quad \textrm{on} \quad 0 < x < \infty,  \label{Eq:NonDimensionalFKPP}
\end{equation}
whose solution $u(x,t)$ can be re-scaled to correspond to particular choices of dimensional parameters $\bar{D}$, $\bar{\lambda}$ and $\bar{K}$ for a particular application.

Relevant boundary conditions are $\partial u / \partial x = 0$ at $x=0$ and $u \to 0$ as $x \to \infty$.  For convenience, all time-dependent PDE problems on $0 < x < \infty$ considered in this review focus on initial distributions of $u$ that are concentrated near the origin.  To generate numerical solutions of Equation (\ref{Eq:NonDimensionalFKPP}) we work on a truncated domain,  $0 < x < L$, where $L$ is chosen to be sufficiently large that we see clear numerical evidence of travelling wave solutions within a sufficiently long, but necessarily finite duration of time.  For this study, we consider the initial condition
\begin{equation}
u(x,0)=
    \begin{cases}
        u_0 & \text{if } x < b, \\
        u_0 \, \textrm{e}^{-a(x - b)} & \text{if } x > b,
    \end{cases}\label{eq:FisherInitialCondition}
\end{equation}
where $a >0$ is the exponential decay rate of the initial density, $u_0> 0$ is the initial maximum population density, and $b > 0$ is an initial domain length where $u(x,0) = u_0$.  For computational purposes we specify no flux boundary conditions at $x=L$, noting that if $L$ is chosen to be sufficiently large the boundary condition at $x=L$ does not play an important role in the time-dependent solution.

To explore the time-dependent solution of Equation (\ref{Eq:NonDimensionalFKPP}) we discretize $0 < x < L$ uniformly with mesh spacing $\delta x$ so that the location of the $i$th mesh point is given by $x_i = (i-1)\delta x$ for $i=1,2,3,\ldots,N$.  For simplicity, we will write $u(x_i,t)$ as $u_i$ to represent the density at the $i$th mesh point at time $t$.  Taking a method-of-lines approach~\cite{Kreyszig2006,Morton2011}, we approximate the spatial derivative terms in Equation (\ref{Eq:NonDimensionalFKPP}) using a central difference approximation to give a system of semi-discrete coupled nonlinear ordinary differential equations (ODE),
\begin{equation}
\dfrac{\textrm{d}u_i}{\textrm{d}t}=
    \begin{cases}
        \dfrac{u_{i+1}-u_i}{(\delta x)^2} + u_i (1 - u_i), & \text{if } i=1, \\
        \dfrac{u_{i-1} - 2u_i + u_{i+1}}{(\delta x)^2} +  u_i (1 - u_i),    & \text{if } 2 \le i \le N-1, \\
        \dfrac{u_{i-1} - u_i}{(\delta x)^2} + u_i (1 - u_i),  & \text{if } i = N.
    \end{cases} \label{eq:FisherSemiDiscreteSystem}
\end{equation}
Evaluating (\ref{eq:FisherInitialCondition}) on the finite difference mesh gives the initial condition for the solution of the semi-discrete system (\ref{eq:FisherSemiDiscreteSystem}).    This system of $N$ ODEs can be efficiently integrated through time to give $u_i$ for $i=1,2,3,\ldots,N$ using the DifferentialEquations.jl package in Julia~\cite{Rackauckas2017}.  Using standard convergence tolerance options, this package implements a range of numerical methods to incorporate automatic time stepping and truncation error control; for this work we implement a standard 4--5 Runge-Kutta approximation~\cite{Tsitouras2011}.

\begin{figure}[htp]
     \includegraphics[width=1.0\textwidth]{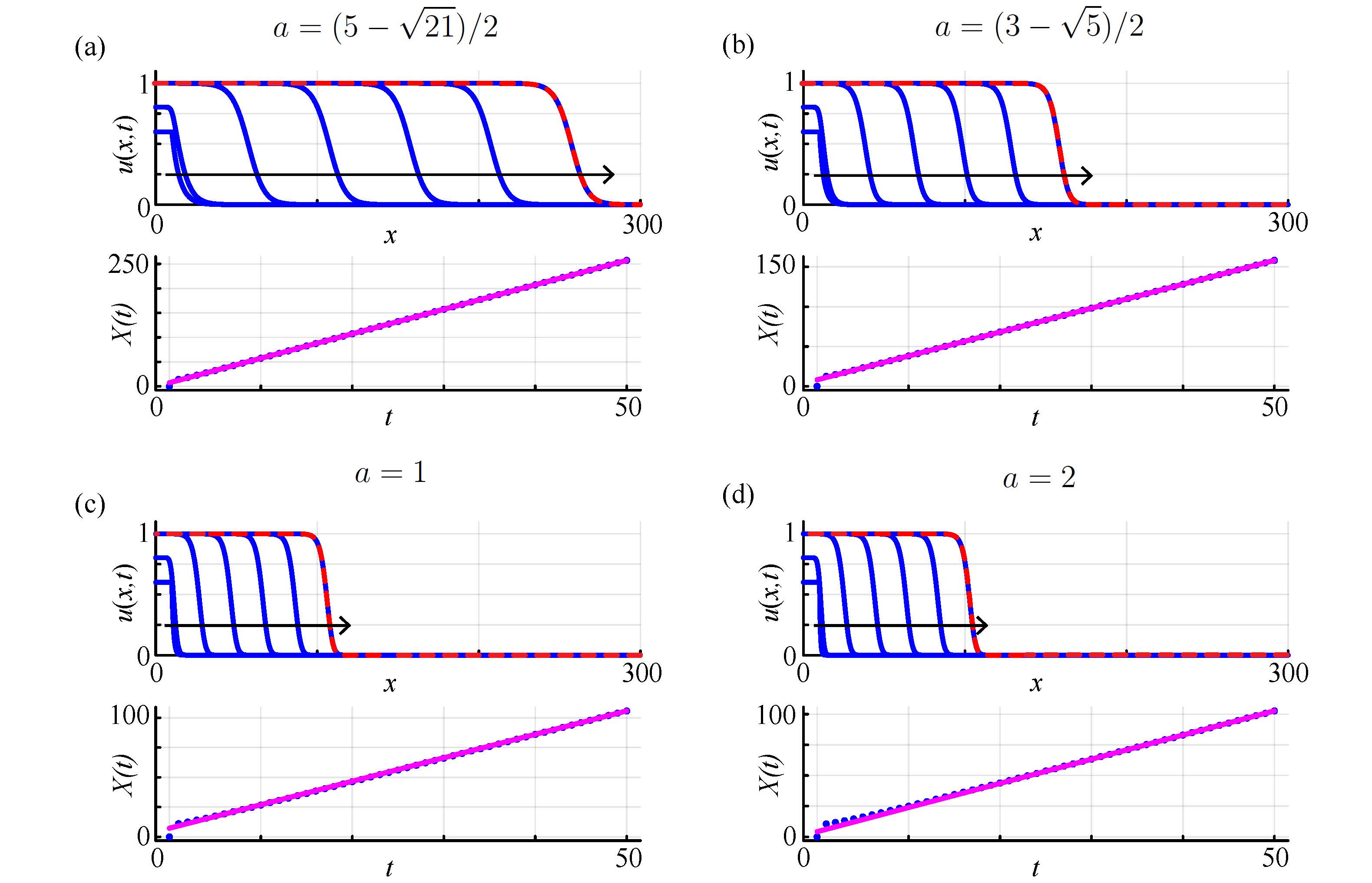}
\caption{Time-dependent solutions of the Fisher-KPP model illustrate how different initial decay rates of $u(x,0)$ affect the long-time travelling wave speed $c$.  Four initial conditions, given by Equation (\ref{eq:FisherInitialCondition}) with $b=10$ and $u_0=0.6$, with varying $a$ are considered: (a) time-dependent PDE solution with $a=(5-\sqrt{21})/2$ leads to $4.999992423$; (b) time-dependent PDE solution with $a=(3-\sqrt{5})/2$ leads to $c=2.999953564$; (b) time-dependent PDE solution with $a=1$ leads to $c=1.970531658$; and, (d) time-dependent PDE solution with $a=2$ leads to $c=1.969077295$.  The upper-panel of each subfigure shows $u(x,t)$ (solid blue) at $t=0,1,10,20,30,40,50$ with the arrow showing the direction of increasing time.   Each time-dependent PDE solution at $t=50$ is superimposed with a dashed red line which corresponds to the shifted $U(z)$ profile obtained in the phase plane. The lower panel shows the time evolution of $X(t)$ where $u(X,t)=0.5$ for $t=0,1,2,3, \ldots, 50$ (blue dots). Front position data in the lower panel of each subfigure is superimposed with a straight line regression (solid pink) obtained using the last five time points.  The slope of the straight line regression gives an estimate of the long-time travelling wave speed, $c$.  Results obtained on $0 \le x \le 300$ with $\delta x = 0.1$.} \label{F2}
\end{figure}

Before presenting time-dependent solutions of the Fisher-KPP model, it is worth noting that the only parameters we specify are the initial decay rate $a$, the initial length of the domain occupied by the density profile $b$, and the initial maximum density $u_0$. Preliminary numerical results (not shown) support the well-known results that the long-time travelling wave solutions are independent of $u_0$ and $b$, so we focus on the impact of different choices of $a$ for a fixed choice of $b$ and $u_0$. Results in Figure \ref{F2} show four different time-dependent solutions with different values of $a$;  we will explain how we chose these particular values later.   Results in the upper panel of Figure \ref{F2}(a) correspond to $a=(5-\sqrt{21})/2$ where we show $u(x,t)$ at $t=0, 1, 10, 20, 30, 40, 50$ with the arrow showing the direction of increasing time.  The initial condition shows the initial decay of $u(x,0)$ with position and the short-time solution $u(x,1)$ shows the impact of the logistic growth term as the density near the origin increases.  The subsequent time-dependent solutions at $t=10, 20, 30, 40, 50$ illustrates that the time-dependent solution approaches a constant shape wave that appears to move with constant speed.  To quantify the speed of the wave, we store the solution $u(x,t)$ at $t=0, 1,2, \ldots, 50$, and at each of the 51 time points we use linear interpolation to compute the value of $X$ so that $u(X,t)= 1/2$.  This computational procedure implicitly defines a function, $X(t)$, describing the time evolution of the position of the wave front.  The lower panel of each subfigure in Figure \ref{F2} gives a scatter plot of $X(t)$ for $t=0,1,2,3, \ldots, 50$ where we see that these points settle to a straight line with positive slope that represents the speed of the wave front during the simulation period.  We quantify the speed of the front by taking late-time data, here chosen to be the last five time points (i.e. $t=46, 47, 48, 49, 50$) and fitting a straight line to this data to give an estimate of the speed of the front, $c$.

As suggested by the computational results in Figure \ref{F2}, the long-time travelling wave speed is related to the spatial decay rate of the initial condition, $a$.  It is possible to formalise this putative relationship by supposing we have $u(x,0) \sim u_0 \textrm{exp}(-ax)$ as $x \to \infty$, and considering the motion of the leading edge where $u \ll 1$.   Under these conditions the evolution of the leading edge can be approximated by the solution of a linear PDE~\cite{Skellam1951} by writing $u \sim \hat{u}$, so that
\begin{equation}
\dfrac{\partial \hat{u}}{\partial t} =  \dfrac{\partial^2 \hat{u}}{\partial x^2} +  \hat{u},  \quad \textrm{for} \quad  \hat{u} \ll 1.  \label{eq:LinearisedFKPP}
\end{equation}
Assuming the long-time solution takes the form of a constant speed travelling wave with an exponentially decaying front, $\hat{u}(x,t)=\textrm{const}\textrm{exp}(-a(x-ct))$, where $c$ is the travelling wave speed, we substitute this solution into the linearised PDE to give
\begin{equation}
c = a + \dfrac{1}{a}, \label{eq:Dispersion}
\end{equation}
which relates the decay rate of the initial condition $a$ to the long-time travelling wave speed, $c$.    This expression is consistent with the results in Figure \ref{F2}(a)--(c), however the computational results in Figure \ref{F2}(d) lead to a slower observed travelling wave speed that indicated by Equation (\ref{eq:Dispersion}).  The reason for the slow than expected travelling wave in Figure \ref{F2}(d) is that for $a > 1$, a comparison principle limits travelling wave speeds to be  $c \le 2$~\cite{Larson1978}.  Furthermore, a travelling wave analysis (below) shows that the minimum travelling wave speed possible is $c=2$,  Therefore, for $a > 1$ we have $c=2$.   Together, these conditions give a continuous piecewise relationship
\begin{equation}
c =
    \begin{cases}
       a + \dfrac{1}{a},& \text{if } a <  1, \\
       2, & \text{if } a > 1, \label{eq:FisherDispersion}
    \end{cases}
\end{equation}
where for  $a$ sufficiently large the long-time travelling wave speed is independent of the initial decay rate.  This explains the results in Figure \ref{F2} where we see that slowly decaying initial conditions lead to travelling wave solutions with $c > 2$, whereas rapidly decaying initial conditions with $a \ge 1$ lead to travelling wave solutions that eventually move with the same minimum wave speed $c = 2$.  Equation (\ref{eq:FisherDispersion}) is often called the \textit{dispersion relationship}.

To provide more insight into travelling wave solutions of the Fisher-KPP model, we explore the phase plane by writing the long time travelling wave solution as $u(x,t) = U(z)$, where $z = x - ct$.  It is important to emphasise that working in the travelling wave coordinate implicitly assumes that we are dealing with a long-time solution of the time-dependent PDE where a constant speed and constant shape travelling wave solution has developed.  Under these conditions, transforming the time-dependent PDE model into the travelling wave coordinate gives
\begin{equation}
\dfrac{\textrm{d}^2U}{\textrm{d}z^2}+ c \dfrac{\textrm{d}U}{\textrm{d}z} + U(1-U)=0, \quad \textrm{on} \quad -\infty < z < \infty, \label{eq:FKPPTravellingWave}
\end{equation}
with boundary conditions
\begin{equation}
\lim_{z \to \infty}U(z) = 0, \quad \textrm{and} \quad  \lim_{z \to -\infty}U(z) = 1. \label{eq:FKPPTravellingWaveBC}
\end{equation}
Note that (\ref{eq:FKPPTravellingWave})--(\ref{eq:FKPPTravellingWaveBC}) is invariant under translation in $z$, which means that we may fix the solution by setting $U(0)=1/2$, for example. Exact, closed-form solutions of this boundary value problem for $U(z)$ are unknown except for the special case of $c^2 = 25/6$ where (\ref{eq:FKPPTravellingWave}) has the Painlev{\'e} property~\cite{Ablowitz1979,McCue2021}.

To make progress, we study the solution of (\ref{eq:FKPPTravellingWave})--(\ref{eq:FKPPTravellingWaveBC}) in the $(U,V)$ phase plane, where
\begin{align}
\dfrac{\textrm{d}U}{\textrm{d}z} &= V, \label{eq:PhasePlane DS1}\\
\dfrac{\textrm{d}V}{\textrm{d}z} &= -cV - U(1-U). \label{eq:PhasePlane DS2}
\end{align}
At this point it is interesting to comment on the differences between studying travelling wave solutions of the Fisher-KPP model using the phase plane as opposed to considering long-time solutions of the time-dependent PDE model.  As we pointed out previously, the latter involves specifying details of the initial condition, such as the decay rate $a$, and then solving the PDE to estimate the long-time travelling wave speed $c$ as an \textit{output} of this computational task.  In contrast, working with the phase plane implicitly assumes that we have moved to the long--time limit where a travelling wave solution $U(z)$ is relevant, which involves specifying the travelling wave speed $c$ as an \textit{input} to this separate computational task.  This means that we can visualise the phase plane for any value of $c$ that we choose, regardless of whether that value arises as a long--time output of a time--dependent PDE solution. This includes travelling wave solutions that are physically unrealistic, as we will explain and demonstrate later.

The dynamical system governing the phase plane (\ref{eq:PhasePlane DS1})--(\ref{eq:PhasePlane DS2}) has two equilibrium points: $(\bar{U},\bar{V})=(0,0)$ and $(\bar{U},\bar{V})=(1,0)$.  Linear stability indicates that $(\bar{U},\bar{V})=(1,0)$ is a saddle point for all values of $c$.  The unstable and stable manifolds near the saddle point tend to straight lines passing through $(\bar{U},\bar{V})=(1,0)$ with slopes given by
\begin{equation}
m_{\pm}  = \dfrac{-c \pm \sqrt{c^2 + 4}}{2}.
\end{equation}
Linear stability analysis indicates that  $(\bar{U},\bar{V})=(0,0)$ is a stable spiral for $c < 2$ and a stable node for $c \ge 2$.  As we shall explain, this bifurcation at $c=2$ shows that all physically realistic travelling wave solution to the Fisher-KPP model have $c \ge 2$.

\begin{figure}[htp]
     \includegraphics[width=1.0\textwidth]{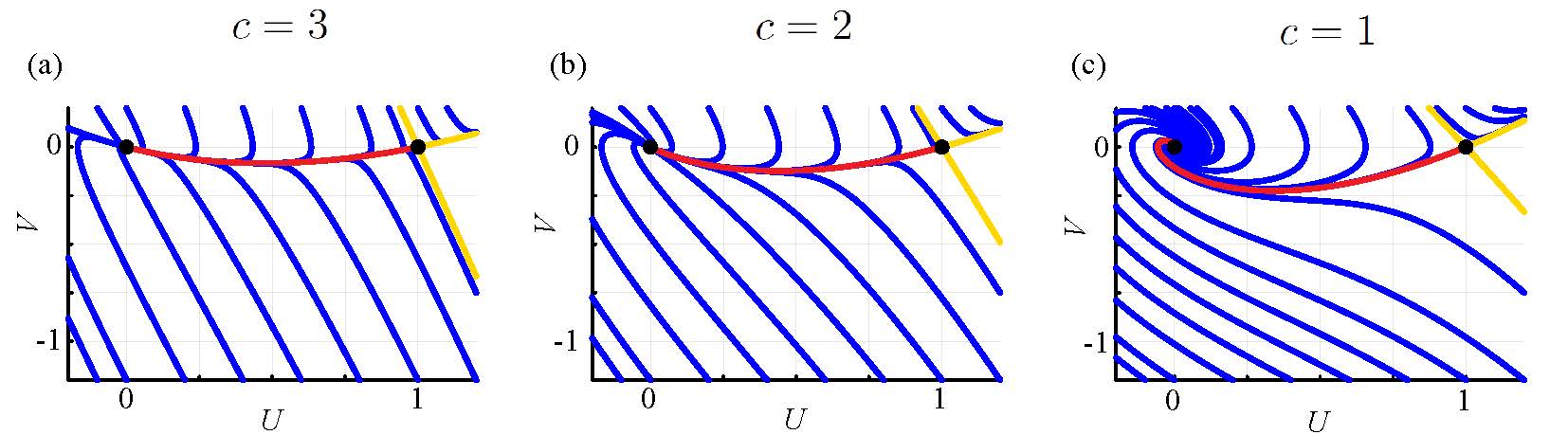}
\caption{Phase planes for (\ref{eq:PhasePlane DS1})--(\ref{eq:PhasePlane DS2}) with: (a) $c=3$; (b) $c=2$; and (c) $c=1$.  Each phase plane show the location of the equilibria (black discs) and several solutions of  (\ref{eq:PhasePlane DS1})--(\ref{eq:PhasePlane DS2}).  Several solutions (solid blue) are included to provide a complete picture of the dynamical system.  Solutions in yellow correspond to those solutions that enter or leave $(1,0)$ along the stable and unstable manifolds.  The heteroclinic orbit leaving $(1,0)$ along the unstable manifold and entering the fourth quadrant before terminating at the origin is shown in red.  Comparing the trajectories near the origin confirm that the origin is a stable node for $c  \ge 2$ and a stable spiral for $c < 2$.} \label{F3}
\end{figure}

With this information, we can now visualise various phase planes and explore the consequences of varying $c$.  Results in Figure \ref{F3} illustrate key features of the phase planes for $c=3,2,1$, respectively.  Trajectories in the phase planes are obtained by solving (\ref{eq:PhasePlane DS1})--(\ref{eq:PhasePlane DS2}) numerically using DifferentialEquations.jl in Julia.  The phase plane in Figure \ref{F3}(a) for $c=3$ shows the two equilibria and a range of trajectories in blue that are included to provide visual interpretation  of the full dynamical system.  Each blue trajectory is a valid solution of (\ref{eq:PhasePlane DS1})--(\ref{eq:PhasePlane DS2}), but these solutions are not relevant in the sense that they are not related to travelling wave solutions of the Fisher-KPP model since they do not satisfy the boundary conditions (\ref{eq:FKPPTravellingWaveBC}).  We also show three yellow trajectories and one red trajectory.  These four trajectories enter and leave the equilibrium point at $(1,0)$ along the stable and unstable manifolds and plotting these four trajectories on the phase plane makes the point that $(1,0)$ is a saddle point.  The red trajectory that leaves $(1,0)$ along the unstable manifold into the fourth quadrant of the $(U,V)$ phase plane enters the origin to form  a heteroclinic orbit. Of all the trajectories plotted in the phase plane, it is only this red heteroclinic orbit that has a physical interpretation in terms of the travelling wave solution since it satisfies the boundary conditions (\ref{eq:FKPPTravellingWaveBC}).

We can interpret this red heteroclinic orbit in the phase plane as a parametric curve $(U(z), V(z))$ from which we can plot $U(z)$ and compare the shape of this curve, after applying an appropriate shift in $z$, with a long-time solution of the time-dependent PDE.  The dashed red curve in Figure \ref{F2}(b) corresponds to $U(z)$ generated in the phase plane, appropriately shifted and superimposed
on the late-time PDE solution $u(x,50)$.  Comparing these two curves shows that the shape of the travelling wave solution obtained from the phase plane is visually indistinguishable from the shape of the late-time PDE solution on this scale.  Indeed, each upper panel in the four subplots in Figure \ref{F2} compares the shape of a late-time PDE solution with the shape of $U(z)$, appropriately shifted, obtained from the phase plane for the corresponding value of $c$.

Results in Figure \ref{F3}(b) shows a similar phase plane, together with many illustrative blue solution trajectories and the three yellow and one red solution trajectory for $c=2$.  Linear stability indicates that the solution of (\ref{eq:PhasePlane DS1})--(\ref{eq:PhasePlane DS2}) locally near the origin bifurcates at $c=2$ since the origin is a stable node for $c \ge 2$ and a stable spiral for $c < 2$.  The borderline case of $c=2$, shown in Figure \ref{F3}(b) indicates that the heteroclinic orbit enters the origin with $U > 0$ as $z \to \infty$.  Taking the shape of the $U(z)$ solution from the phase plane, shifting this profile and superimposing on the late-time PDE solutions in Figure \ref{F2}(c)--(d) shows that the shape of the wave obtained from the phase plane is visually indistinguishable from these two late-time PDE solutions that evolved from different initial conditions.

The phase plane in Figure \ref{F2}(c) for $c=1$ shows similar features to the two previous phase planes with the exception that we now see the solution of (\ref{eq:PhasePlane DS1})--(\ref{eq:PhasePlane DS2}) spirals into the origin, which implies that $U < 0$ over infinitely many subintervals for  $z > 0$.  While this trajectory is a valid solution of the dynamical system together with the boundary conditions (\ref{eq:FKPPTravellingWaveBC}), the heteroclinic orbit is not physical in the sense that it does not relate to a late-time solution of the Fisher-KPP model.  Therefore, we see that from phase-plane analysis alone, we must have $c \ge 2$. Furthermore, this example makes the point we alluded to earlier that working in the phase plane allows us to specify a value of $c$ regardless of whether that value relates to a long-time output of a time-dependent PDE solution.  This observation has often led researchers to discard this phase plane for $c < 2$, however we will revisit this question later in Section \ref{sec:FisherStefan} when we consider the Fisher-Stefan model.

\begin{figure}[htp]
     \includegraphics[width=1.0\textwidth]{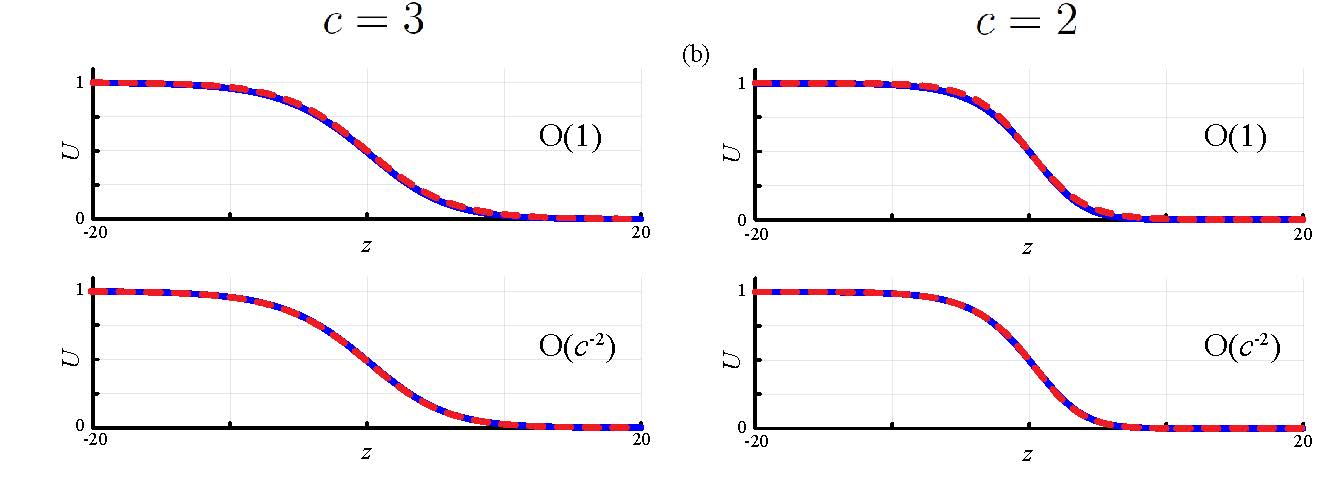}
\caption{Approximate travelling wave shape given by $c \to \infty$ asymptotic expressions for: (a) $c=3$, and (b) $c=2$. All approximate solutions are compared with the shape of the travelling wave obtained by taking a late-time PDE solution from Figure \ref{F2}(b)-(d), translates so that $U(0)=1/2$ (solid blue).  The upper panels compares the full numerical solution with the  $\mathcal{O}(1)$ perturbation approximation $U0(z)$ and the lower panel compares the full numerical solution with the $\mathcal{O}(c^2)$ perturbation approximation $U_0(z) + c^{-2}U_1(z)$.} \label{F4}
\end{figure}

Computational explorations of the time-dependent solutions of the Fisher-KPP model, along with our leading edge and phase plane analysis confirm that travelling wave solutions of the Fisher-KPP model involve $c \ge 2$. We will now show that it is possible to develop some mathematical insight into the shape of these travelling waves in the limit $c \to \infty$ by following Canosa~\cite{Canosa1973}.  Introducing the change of coordinates $\xi = z / c$ allows us to re-write Equation (\ref{eq:FKPPTravellingWave}) as
\begin{equation}
\dfrac{1}{c^2} \dfrac{\textrm{d}^2U}{\textrm{d}\xi^2}+  \dfrac{\textrm{d}U}{\textrm{d}\xi} + U(1-U)=0,  \quad \textrm{on} \quad -\infty < z < \infty, \label{eq:CanosaFisherKPPBVP}
\end{equation}
with boundary conditions
\begin{equation}
\lim_{\xi \to \infty}U(\xi) = 0, \quad \textrm{and} \quad  \lim_{\xi \to -\infty}U(\xi) = 1.
\end{equation}
Identifying the small parameter $c^{-1}$ and assuming the solution can be written as a regular perturbation expansion~\cite{Murray1984}
\begin{equation}
U(\xi) = U_0(\xi) + \left(\dfrac{1}{c^2}\right)U_1(\xi) + \mathcal{O}\left(\dfrac{1}{c^4}\right),
\end{equation}
we substitute this expansion into (\ref{eq:CanosaFisherKPPBVP}) to give
\begin{align}
\dfrac{\textrm{d}U_0}{\textrm{d}\xi} &= -U_0(1-U_0), \quad U_0(0) = \dfrac{1}{2},\\
\dfrac{\textrm{d}U_1}{\textrm{d}\xi} &= -\dfrac{\textrm{d}^2U_0}{\textrm{d}\xi^2} - U_1(1-2U_0),    \quad U_1(0) = 0,
\end{align}
whose solutions, $U_0(\xi)$ and $U_1(\xi)$, can be written in terms of the original independent variable as
\begin{align}
U_0(z) &= \dfrac{1}{1+\textrm{e}^{z/c}},\\
U_1(z) &= \dfrac{\textrm{e}^{z/c}}{\left(1+\textrm{e}^{z/c}\right)^2}\ln\left[\dfrac{4 \textrm{e}^{z/c} }{\left(1 + \textrm{e}^{z/c}\right)^2} \right].
\end{align}
With these expressions for $U_0(z)$ and $U_1(z)$, we now have a two-term perturbation solution $U(z) = U_0(z) + c^{-2}U_1(z)$ that provides insight into how the shape of the travelling wave solution relates to the travelling wave speed $c$.  Profiles in Figure \ref{F4}(a) compare the leading order solution $U_0(z)$, and the two-term perturbation solution  $U_0(z) + c^{-2}U_1(z)$ with a late-time solution of the time-dependent PDE for $c=3$.  In this case we see that the leading order solution provides a very good approximation of the full travelling wave solution since the differences between the numerical result and the perturbation solution barely evident at this scale.  Further, we see that the two-term perturbation solution is visually indistinguishable from the numerical solution at this scale.  Profiles in Figure \ref{F4}(b) are for the minimum travelling wave speed, $c=2$, where again we see that the two-term perturbation solution accurately matches the numerical result.  The fact that the perturbation solution performs very well for the minimum travelling wave speed illustrates how useful perturbation solutions can be as we are unable to obtain closed form exact solutions for arbitrary values of $c$, yet these approximate solutions valid in the limit that $c \to \infty$ turn out to be relatively accurate for just $c=2$.

More generally, our perturbation solution provides insight into the relationship between the shape of the travelling wave solution and the speed of the wave, $c$.  For example, evaluating the $\mathcal{O}(1)$ perturbation solution at $z=0$ gives
\begin{equation}
\dfrac{\textrm{d} U}{\textrm{d}z} = -\dfrac{1}{4c} + \mathcal{O}\left(\dfrac{1}{c^2}\right),
\end{equation}
indicating that the faster the travelling wave speed, the flatter the shape of the wave since we have $\textrm{d}U/\textrm{d}z \to 0^{-}$ as $c \to \infty$~\cite{Murray2002}.

\section{Porous-Fisher model: Smooth initial conditions}\label{sec:PorousFisherSmooth}
As explained in Section \ref{sec:FisherSmooth}, one of the limitations of travelling wave solutions of the Fisher-KPP model is that they are smooth with  $u > 0$ for all $x > 0$, which means that it is difficult to unambiguously identify the precise position of the leading edge of the moving front.  This is a challenge because experimental observations often lead to clearly defined sharp fronts that are incompatible with this feature of the Fisher-KPP model~\cite{Maini2004a,Maini2004b,McCue2019}.  As such, considerable effort has been spent in developing and understanding alternative reaction-diffusion models that lead to travelling wave solutions with compact support so that these solutions can be used to clearly identify the position of the moving front.  Possibly the most common approach is to work with the Porous-Fisher model, which can be written in non-dimensional form as
\begin{equation}
\dfrac{\partial u}{\partial t} = \dfrac{\partial}{\partial x}\left[u\dfrac{\partial u}{\partial x}\right]  + u(1-u),  \quad \textrm{on} \quad 0 < x < \infty, \label{eq:PorousFisherPDE}
\end{equation}
with $\partial u / \partial x = 0$ at $x=0$ and $u \to 0$ as $x \to \infty$.  The key difference is that the linear diffusion flux in the Fisher-KPP model $\mathcal{J} = -\partial u / \partial x$ is replaced with a degenerate nonlinear diffusive flux $\mathcal{J} = -u \, \partial u / \partial x$ so that $\mathcal{J}$ vanishes  when $u=0$~\cite{Pattle1959,Vazquez2006,Simpson2024c}.  For numerical purposes, we will replace the far-field condition with $\partial u / \partial x = 0$ at $x=L$, where the value of $L$ is chosen to be sufficiently large so that this does not affect our numerical solution.   We will consider the same initial condition as for the Fisher-KPP model, namely (\ref{eq:FisherInitialCondition}).    To explore time-dependent solutions of the Porous-Fisher model we discretise Equation (\ref{eq:PorousFisherPDE}) on the same equally-spaced finite difference mesh to give,
\begin{equation}
\dfrac{\textrm{d}u_i}{\textrm{d}t}=
    \begin{cases}
        \dfrac{(u_{i+1}+u_{i})(u_{i+1}-u_{i})}{2(\delta x)^2} +u_i (1 - u_i), & \text{if } i=1, \\
        \dfrac{(u_{i+1}+u_{i})(u_{i+1}-u_{i}) - (u_{i-1}+u_{i})(u_{i}-u_{i-1})}{2(\delta x)^2} + u_i (1 - u_i),    & \text{if } 2 \le i \le N-1, \\
        \dfrac{(u_{i-1}+u_{i})(u_{i-1}-u_{i})}{2(\delta x)^2} + u_i (1 - u_i), & \text{if } i = N,
    \end{cases}
\end{equation}
which can be integrated through time using the same Julia package as for the Fisher-KPP model.  The inputs into these time-dependent PDE solutions are the parameters governing the initial condition $u_0$, $a$ and $b$; long-time PDE solutions can be used to estimate the travelling wave speed, $c$.  Again, as for the Fisher-KPP model, preliminary explorations (now shown) indicate that  $c$ depends upon $a$, but is independent of $u_0$ and $b$.

\begin{figure}[htp]
\centering
     \includegraphics[width=1.0\textwidth]{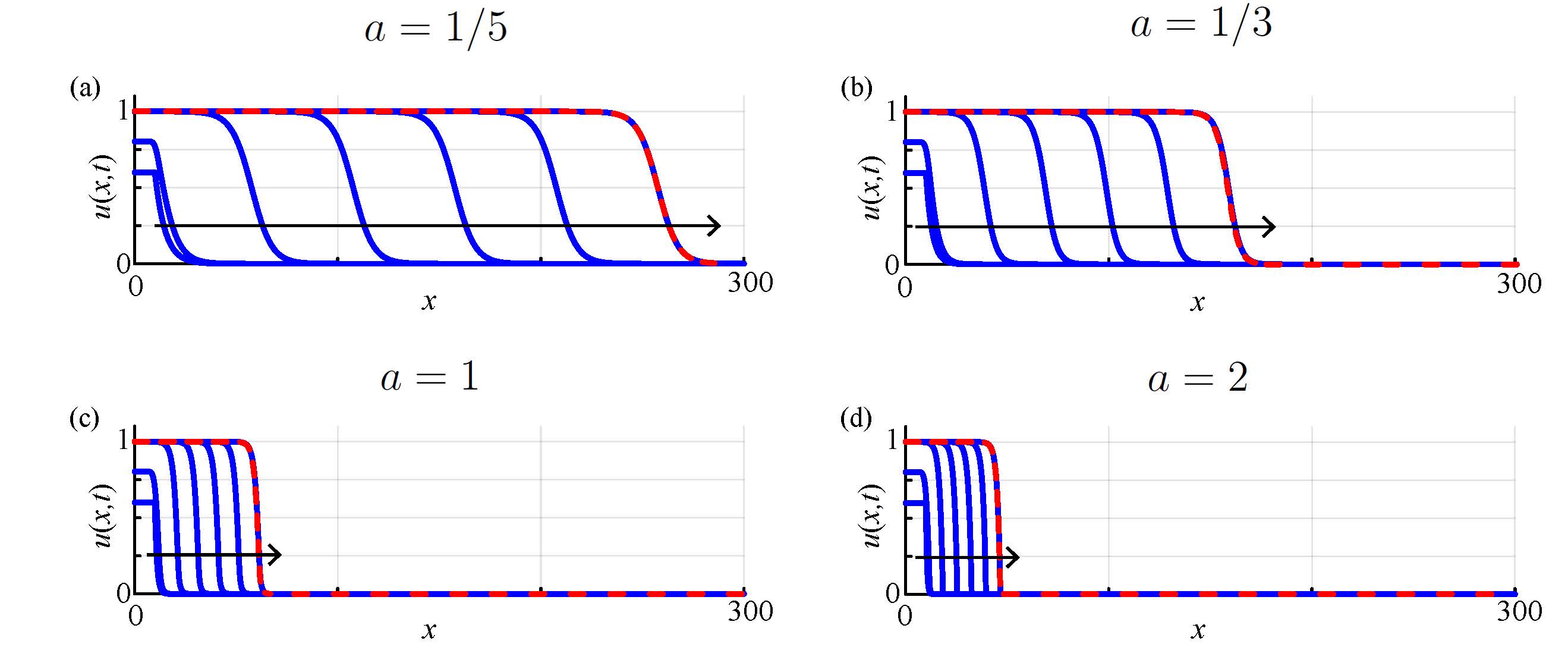}
\renewcommand{\baselinestretch}{1.0}
\caption{Time-dependent solutions of the Porous-Fisher model illustrate how different initial decay rates of $u(x,0)$ affect the long-time travelling wave speed $c$.  Four initial conditions, given by Equation (\ref{eq:FisherInitialCondition}) with $u_0=0.6$ and $b=10$, with varying $a$ are considered: (a) time-dependent PDE solution with $a=1/5$ leads to $5.000000000$; (b) time-dependent PDE solution with $a=1/3$ leads to $c=3.000000000$ (b) time-dependent PDE solution with $a=1$ leads to $c=0.999999999$; and, (d) time-dependent PDE solution with $a=2$ leads to $c=0.704887285$.   Each time dependent PDE solution at $t=50$ is superimposed with a dashed red line which corresponds to the shifted $U(z)$ profile obtained in the phase plane. Results obtained on $0 \le x \le 300$ with $\delta x = 0.02$.}\label{F5}
\end{figure}

Results in Figure \ref{F5} illustrate time-dependent solutions of the Porous-Fisher model with the initial condition given by Equation (\ref{eq:FisherInitialCondition}) with $a=1/5, 1/3, 1, 2$.  In each case, we see the eventual formation of a constant speed, constant shape travelling wave.  Using the same approach as for the Fisher-KPP model, we trace the evolution of $X(t)$, where $u(X,t)=0.5$, to generate a time series of $X(t)$ data which eventually falls on a straight line with a positive slope.  Fitting a straight line regression model to the last five points in this time series gives us an estimate of $c$.  Details of the $X(t)$ data and the straight line regression plot are not shown here as they look almost identical to those shown in Figure \ref{F2} for the Fisher-KPP model.  More information can be obtained by exploring the time-dependent PDE codes available on  \href{https://github.com/ProfMJSimpson/PDEInvasion}{GitHub}.

An important point illustrated by the time-dependent PDE solutions in Figure \ref{F5} is that solutions with sufficiently slow initial decay rates lead to fast, smooth-fronted travelling wave solutions that qualitatively resemble travelling wave solutions to the Fisher-KPP model in Figure \ref{F2}.  Later in this Section we will provide a mathematical explanation of this observation.  On the other hand, highlighted in Figure \ref{F5}(d) is that the smooth initial condition with a sufficiently fast exponential decay appears to evolve into a sharp-fronted travelling wave with a well-defined front position.  This is a very different outcome to the Fisher-KPP case and may be interpreted as a favourable feature in terms of mathematical modelling.  We will investigate the sharp-fronted solution later when we explore these travelling wave solutions in the phase plane, and then provide a more detailed explanation of these results in Section \ref{Sec:Compact}.

From a numerical point of view it is worth noting that results in Figures \ref{F2} and \ref{F3} are obtained on the same domain, $0 \le x \le 300$, but we have used a much finer discretisation to explore time-dependent solutions of the Porous-Fisher model.  Our explorations (not shown) indicate that smooth-fronted solutions of the Porous-Fisher model and the Fisher-KPP model can be obtained using a relatively coarse discretisation (e.g. $\delta x = 0.1$) leading to grid-independent results when plotted on this scale.  The situation is different, however, with the apparent sharp-fronted solution of the Porous-Fisher model in Figure \ref{F3}(d) where we require a much finer numerical discretisation to accurately resolve the shape of the leading edge of the travelling wave, and to provide an accurate estimate of $c$.  Therefore, unlike the Fisher-KPP model where we find that it is possible to make an appropriate choice of $\delta x$ so that the numerical algorithm performs reasonably well for a range of $a$ values, much greater care needs to be exercised when solving the Porous-Fisher model numerically because the possibility of sharp-fronted solutions requires much finer resolution.  For simplicity, all results in Figure \ref{F5} are generated with a much finer mesh resolution than the results in Figure \ref{F2} for the Fisher-KPP model. We encourage the reader to explore the impact of choosing different values of $\delta x$ using the code provided as it is straightforward to show that smooth travelling wave solutions of the Porous-Fisher model can be very accurate with a much coarser mesh.

Just as we did for the smooth-fronted travelling wave solutions of the Fisher-KPP model, it is possible to relate the long-time travelling wave speed $c$ to the decay rate of the initial condition for smooth-fronted travelling wave solutions of the Porous-Fisher model.  Supposing we have $u(x,0) \sim u_0\textrm{exp}(-ax)$ as $x \to \infty$ for $a>0$, we can examine the motion of the leading edge by writing $u \sim \hat{u}$ so that
\begin{equation}
\dfrac{\partial \hat{u}}{\partial t} =  \hat{u},  \quad \textrm{for} \quad \hat{u} \ll 1. \label{eq:PorousFisherLinearisedPDE}
\end{equation}
If the travelling wave solution of (\ref{eq:PorousFisherPDE}) takes the form $\hat{u}(x,t) =\textrm{const} \, \textrm{exp}(-a(x-ct))$, substituting this solution into the linearised PDE (\ref{eq:PorousFisherLinearisedPDE}) gives the dispersion relationship for the Porous-Fisher model
\begin{equation}
c = \dfrac{1}{a},
\end{equation}
which is consistent with the smooth-fronted travelling wave solutions presented in Figure \ref{F5}(a)--(c).

Unlike the Fisher-KPP model, the minimum travelling wave speed for travelling wave solutions of the Porous-Fisher model cannot be determined solely by examining the dispersion relationship and we now turn to the phase plane by writing the long-time travelling wave solution as $u(x,t) = U(z)$, where $z = x - ct$.  Transforming the time-dependent PDE model into the travelling-wave coordinate gives
\begin{equation}
\dfrac{\textrm{d}}{\textrm{d}z}\left[ U \dfrac{\textrm{d}U}{\textrm{d}z}\right]+ c \dfrac{\textrm{d}U}{\textrm{d}z} + U(1-U) = 0, \quad \textrm{on} \quad -\infty < z < \infty, \label{eq:PhasePlanePF}
\end{equation}
with boundary conditions
\begin{equation}
\lim_{z \to \infty}U(z) = 0, \quad \textrm{and} \quad   \lim_{z \to -\infty}U(z) = 1.
\end{equation}
As for travelling wave solutions of the Fisher-KPP model, it is not obvious how to solve Equation (\ref{eq:PhasePlanePF}) for arbitrary values of $c$, so we study the solution of this boundary value problem in the $(U,V)$ phase plane where we have~\cite{Murray2002}
\begin{align}
\dfrac{\textrm{d}U}{\textrm{d}z} &= V, \\
U\dfrac{\textrm{d}V}{\textrm{d}z} &= -cV - U(1-U)-V^2.
\end{align}
This dynamical system is singular when $U=0$.  The singularity can be removed by introducing a transformed independent variable that we write implicitly as~\cite{Murray2002}
\begin{equation}
\dfrac{\textrm{d} }{\textrm{d} \zeta} = U \dfrac{\textrm{d}}{\textrm{d}z}, \label{eq:Desingualrisedtransform}
\end{equation}
which gives a de-singularised system
\begin{align}
\dfrac{\textrm{d}U}{\textrm{d}\zeta} &= UV, \label{eq:PhasePlane PFDS1} \\
\dfrac{\textrm{d}V}{\textrm{d}\zeta} &= -cV - U(1-U)-V^2. \label{eq:PhasePlane PFDS2}
\end{align}
This desingularised phase plane is has three equilibria: $(\bar{U},\bar{V})=(0,0)$,  $(\bar{U},\bar{V})=(1,0)$ and $(\bar{U},\bar{V})=(0,-c)$.  Linear stability analysis indicates that both $(\bar{U},\bar{V})=(1,0)$ and $(\bar{U},\bar{V})=(0,-c)$ are saddle points for all values of $c$, while the origin is a stable nonlinear node.   Similar to the Fisher-KPP model, smooth travelling waves are associated with a heteroclinic orbit between $(1,0)$ and $(0,0)$, but we now have the possibility of a sharp-fronted heteroclinic orbit between $(1,0)$ and $(0, -c)$.  In this case the  heteroclinic orbit terminates at $(U,V)=(0,-c)$ meaning that the contact point, where $U=0$, has a finite negative slope for $c > 0$.

\begin{figure}[htp]
\centering
     \includegraphics[width=1.0\textwidth]{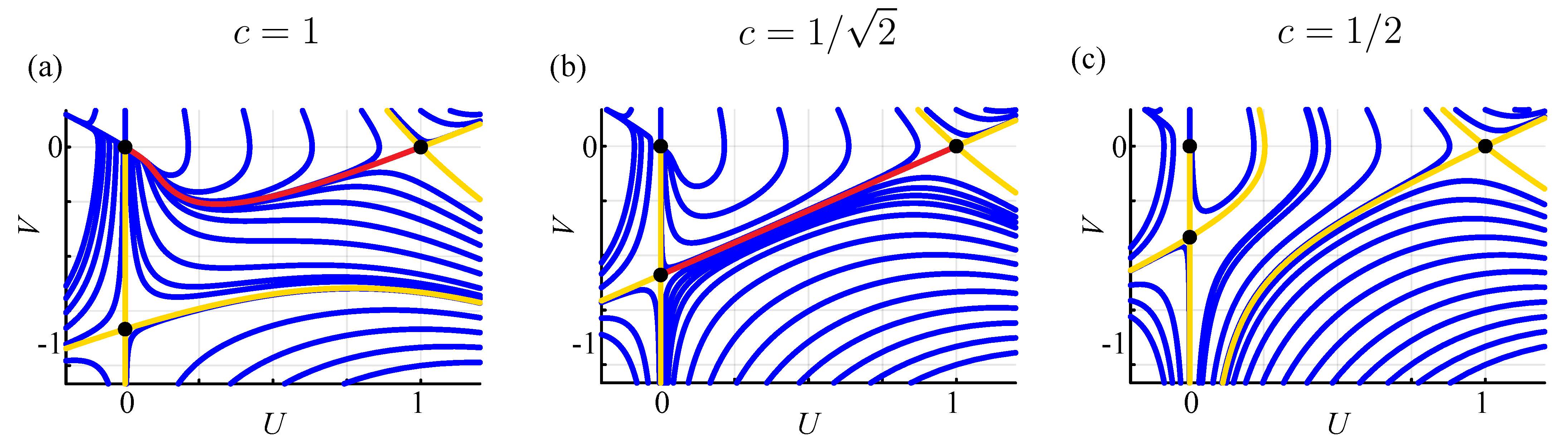}
\renewcommand{\baselinestretch}{1.0}
\caption{Phase planes for (\ref{eq:PhasePlane PFDS1})--(\ref{eq:PhasePlane PFDS2}): (a) $c=1$; (b) $c=1 / \sqrt{2}$; and (c) $c=1/2$.  Each phase plane show the location of the equilibria (black discs) and several trajectories.  Several solutions (solid blue) are included to provide a complete picture of the solution of the dynamical system, solutions in yellow correspond to those solutions that enter or leave $(1,0)$ and $(0,-c)$ along the stable and unstable manifolds.  In (a) the heteroclinic orbit leaving $(1,0)$ along the unstable manifold and entering the fourth quadrant before terminating at the origin is shown in red, corresponding to a smooth-fronted travelling wave solution.  In (b) the straight line heteroclinic orbit leaving $(1,0)$ along the unstable manifold and entering the fourth quadrant before terminating at $(0,-c)$ is shown in red, corresponding to a sharp-fronted travelling wave solution. In (c) there is no heteroclinic orbit between $(1,0)$ and either of the other equilibria, indicating that there is no travelling wave solution for this value of $c$.} \label{F6}
\end{figure}

Computationally-generated phase planes with various trajectories highlighted are given in Figure \ref{F6} for key choices of $c=1,1/\sqrt{2},1/2$.  The phase plane in Figure \ref{F6}(a) for $c=1$  shows a range of blue trajectories obtained by solving (\ref{eq:PhasePlane PFDS1})--(\ref{eq:PhasePlane PFDS2}) numerically using codes provided on \href{https://github.com/ProfMJSimpson/PDEInvasion}{GitHub}.  These blue trajectories provide a visual picture of the phase plane within the vicinity of the three equilibrium points.  The phase plane also includes seven  yellow trajectories that either enter or leave $(0, -c)$ and $(1,0)$ along the associated unstable and stable manifolds of these saddle node points.  Finally, in red we show the heteroclinic orbit leaving $(1,0)$ along the unstable manifold into the fourth quadrant of the $(U,V)$ phase plane, eventually entering the origin.  This heteroclinic orbit corresponds to the smooth travelling wave solution in Figure \ref{F5}(c).

Exploring various phase planes by gradually reducing $c < 1$ indicates that smooth travelling wave solutions, corresponding to a heteroclinic orbit between $(1,0)$ and $(0,0)$, persist until we reach some critical value of $c$ where, as the equilibrium point at $(0, -c)$ moves up the $V$-axis, it becomes possible to have a straight-line heteroclinic orbit between $(1,0)$ and $(0,-c)$, as illustrated in Figure \ref{F6}(b).  The straight-line heteroclinic orbit, $V = -c(1-U)$, can be substituted into (\ref{eq:PhasePlane PFDS1})--(\ref{eq:PhasePlane PFDS2}) to show that the corresponding special travelling wave speed is $c = 1 / \sqrt{2}$~\cite{Murray2002}.  This orbit corresponds to a sharp-fronted travelling wave solution like the results we obtained numerically in Figure \ref{F5}(d).  Since we now have a mathematical expression for the straight line heteroclinic orbit in the phase plane, we can give a closed-form expression for this special sharp-fronted solution
\begin{equation}
U(z)=
    \begin{cases}
        1 - \textrm{e}^{z/\sqrt{2}}, & \text{if } z \le 0, \\
        0,   & \text{if } z > 0, \label{eq:SharpFrontExact}
    \end{cases}
\end{equation}
where this solution has been arbitrarily shifted so the position of the front is $z=0$.  This solution is significant  because, unlike the Fisher-KPP model, the Porous-Fisher model leads to travelling wave solutions with a natural front position meaning that this feature can be used to match with experimental observations more easily than working with a smooth--fronted travelling wave solution~\cite{Sengers2007,Buenzli2020}.  To complete the picture, Figure \ref{F6}(c) shows a phase plane and associated trajectories for $c =1/2 <  1 / \sqrt{2}$ where we now see that there is no longer any possibility of a heteroclinic orbit, indicating that $c= 1/\sqrt{2}$ is the minimum travelling wave speed for the Porous-Fisher model.

For smooth-fronted travelling wave solutions with $c > 1/\sqrt{2}$ we can numerically integrate (\ref{eq:PhasePlane PFDS1})--(\ref{eq:PhasePlane PFDS2}) to give numerical estimates of the trajectory in the $(U,V)$ phase plane.  This trajectory can be interpreted as a parameterised curve with parameter $\zeta$.   Our numerical algorithms on \href{https://github.com/ProfMJSimpson/PDEInvasion}{GitHub} solve for $U(\zeta)$ and $V(\zeta)$, recording values of these quantities at equally-spaced intervals of $\zeta$, with spacing $\delta \zeta$. To compare our numerical estimate of $U(\zeta)$ with numerical solutions of the full time-dependent PDE model in Figure \ref{F5} we use (\ref{eq:Desingualrisedtransform}) to relate $\zeta$ and $z$, noting that
\begin{equation}
z(\zeta) = \int^{\zeta} U(\zeta') \, \textrm{d}\zeta', \label{eq:PorousFisherIntetgral}
\end{equation}
can be evaluated numerically to give a relationship between $z$ and $\zeta$.   Since we store values of $U(\zeta)$ at equally-spaced intervals of $\zeta$, it is straightforward to compute $z(\zeta)$ by applying the trapezoid rule to evaluate the integral in (\ref{eq:PorousFisherIntetgral}).  While other  quadrature methods could be implemented, our approach is simple to implement, and leads to estimates of $U(z)$ that, after appropriate shifting, compare well with various late-time PDE solutions.  Each smooth-fronted time-dependent PDE solution in Figure \ref{F5}(a)--(c) includes a dashed red line superimposed on $u(x,50)$.  These dashed red lines correspond to the appropriately shifted $U(z)$ profile obtained from the phase planes; we see that the shape of the travelling wave solution from the time-dependent PDE solution is visually indistinguishable from the profile obtained using the phase plane trajectories.  The result in Figure \ref{F5}(d) for the sharp-fronted travelling wave moving with the minimum travelling wave speed compares the late time PDE solution with the shifted  exact solution, Equation (\ref{eq:SharpFrontExact}).  Again, as for the smooth-fronted travelling wave solutions, here we see that the time-dependent PDE solutions at late time are visually indistinguishable, at least at this scale, from the exact phase plane result.

To conclude this investigation into travelling wave solutions of the Porous-Fisher model evolving from smooth initial conditions, we now develop approximate perturbation solutions to describe the shape of smooth-fronted travelling wave solutions in the limit $c \to \infty$.  Introducing the change of coordinates, $\xi =  z / c$, gives
\begin{equation}
\dfrac{1}{c^2} \dfrac{\textrm{d}}{\textrm{d}\xi}\left[U\dfrac{\textrm{d}U}{\textrm{d}\xi} \right]+  \dfrac{\textrm{d}U}{\textrm{d}\xi} +  U(1-U)=0, \quad \textrm{on} \quad -\infty < \xi < \infty, \label{eq:PFCanosa}
\end{equation}
with boundary conditions
\begin{equation}
\lim_{\xi \to \infty}U(\xi) = 0, \quad \textrm{and} \quad   \lim_{\xi \to -\infty}U(\xi) = 1.
\end{equation}
Treating $c^{-1}$ as a small parameter, we assume that the solution can be written as
\begin{equation}
U(\xi) = U_0(\xi) + \left(\dfrac{1}{c^2}\right)U_1(\xi) + \mathcal{O}\left(\dfrac{1}{c^4} \right).
\end{equation}
Substituting this expansion into (\ref{eq:PFCanosa}) gives
\begin{align}
\dfrac{\textrm{d}U_0}{\textrm{d}\xi} &= -U_0(1-U_0), \quad U_0(0) = \dfrac{1}{2},\\
\dfrac{\textrm{d}U_1}{\textrm{d}\xi} &= -\dfrac{\textrm{d}}{\textrm{d}\xi}\left[U_0\dfrac{\textrm{d}U_0}{\textrm{d}\xi} \right] - U_1(1-2U_0),    \quad U_1(0) = 0,
\end{align}
whose solutions, $U_0(\xi)$ and $U_1(\xi)$ can be written in terms of the original independent variable as,
\begin{align}
U_0(z) &= \dfrac{1}{1+\textrm{e}^{z/c}},\\
U_1(z) &= \dfrac{\textrm{e}^{z/c}}{\left(1+\textrm{e}^{z/c}\right)^2}
\left[\ln 2-\dfrac{3}{2} + \dfrac{3+\left(1+\textrm{e}^{z/c}\right)\left[z/c -\ln \left(1+\textrm{e}^{z/c} \right) \right]}{1+\textrm{e}^{z/c}} \right].
\end{align}
These expressions give a two-term perturbation solution $U(z) = U_0(z) + c^{-2}U_1(z)$ which provides insight into how the shape of the travelling wave solution relates to $c$, as we will now explore.

\begin{figure}[H]
\centering
     \includegraphics[width=1.0\textwidth]{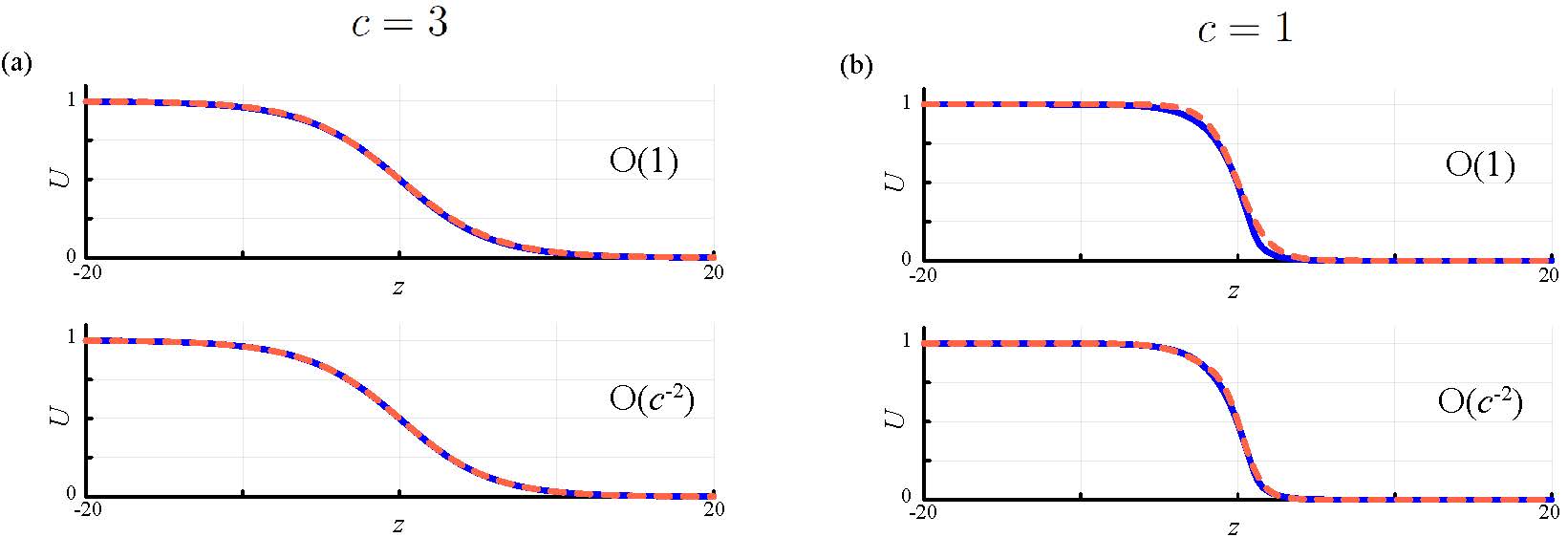}
\renewcommand{\baselinestretch}{1.0}
\caption{Approximate shape of the smooth-fronted travelling wave solution of the PF model given by $c \to \infty$ asymptotic expressions for: (a) $c=3$, and (b) $c=1$. Each approximate solution is compared with the shape of the travelling wave obtained by taking a late-time PDE solution from Figure \ref{F5} translates so that $U(0)=1/2$ (solid blue).  The upper panels compares the full numerical solution with the  $\mathcal{O}(1)$ perturbation approximation $U_0(z)$ and the lower panel compares the full numerical solution with the $\mathcal{O}(c^2)$ perturbation approximation $U_0(z) + c^{-2}U_1(z)$.} \label{F7}
\end{figure}

Profiles in Figure \ref{F7}(a) compare the leading order solution $U_0(z)$, and the two-term perturbation solution  $U_0(z) + c^{-2}U_1(z)$ with a late-time solution of the time-dependent PDE for $c=3$.  As for the Fisher-KPP model, we see that the leading order solution provides a good approximation of the numerical solution of the full problem.  It is useful to recall that travelling wave solutions of the Fisher-KPP model are restricted to $c \ge 2$ and in Figure \ref{F4}(b) we saw that a large $c$ asymptotic approximation performed well, even for $c=2$.  With the Porous-Fisher model we have smooth travelling wave solutions for $c > 1/\sqrt{2}$ which means we can explore the accuracy of a large $c$ asymptotic approximation for even smaller values of $c$ than was possible with the Fisher-KPP model.  Profiles in Figure \ref{F7}(b) for $c=1$ illustrate that the $U_0(z)$ solution is visually-distinct from the full time-dependent solution whereas the two term solution gives a reasonably accurate approximation of the shape of the wave, which is remarkable given that we have $c=1$ and the perturbation approximation is valid for $c \to \infty$.

It is insightful to note that the $U_0(z)$ term in the large $c$ perturbation solution is identical for the Fisher-KPP and Porous-Fisher models. There are several inferences that can be drawn from this observation.  First, we have the same qualitative relationship between the shape and speed of the travelling wave, namely the faster the travelling wave the flatter the shape of the travelling wave for sufficiently large $c$ for both models.  Second, this indicates that the shape of a fast-moving travelling wave is primarily determined by the source term in the PDE model rather than the flux term because the influence of the diffusive flux term arises as a correction term whereas the proliferation mechanism controls the leading order features.   This is biologically insightful because while both diffusion and a source term are required to form a travelling wave solution, it is the proliferation term that largely controls the shape of the travelling wave~\cite{Simpson2006}.

\section{Fisher-KPP and Porous-Fisher models: Initial conditions with compact support}\label{Sec:Compact}
All time-dependent PDE solutions considered in Sections \ref{sec:FisherSmooth}--\ref{sec:PorousFisherSmooth} are mathematically convenient since they have $u(x,0) \sim u_0 \, \textrm{exp}(-ax)$ as $x \to \infty$ for $a > 0$.  This means that the initial conditions shown in Figures \ref{F2} and \ref{F5} are smooth with $u(x,0) > 0$ for all $x > 0$.  This form of initial condition allows us derive a dispersion relationship that relates the exponential decay rate $a$ to the long-time travelling wave speed $c$, and our time-dependent PDE solutions in Figures \ref{F2} and \ref{F5} are consistent with the dispersion relationship in both cases.  This approach, however, is not always preferred since real applications of biological invasion are unlikely to involve an initial distribution that decays exponentially with position~\cite{Jin2016,Steel1998}.  An alternative approach is to consider an initial condition of the form
\begin{equation}
u(x,0)=
    \begin{cases}
        u_0, & \text{if } 0 < x < b, \\
        0,   & \text{if } x > b, \label{eq:FisherICCompactSupport}
    \end{cases}
\end{equation}
which is far more practical because it models some initial region that is occupied at density $u_0$, that is adjacent to another region, $x > b$, which is completely vacant with $u=0$.   This is precisely the kind of initial condition that is used in the case of applications in cell biology, such as modelling scratch assay experiments~\cite{Sengers2007,Simpson2006,Maini2004a,Maini2004b}.

\begin{figure}[htp]
\centering
     \includegraphics[width=1.0\textwidth]{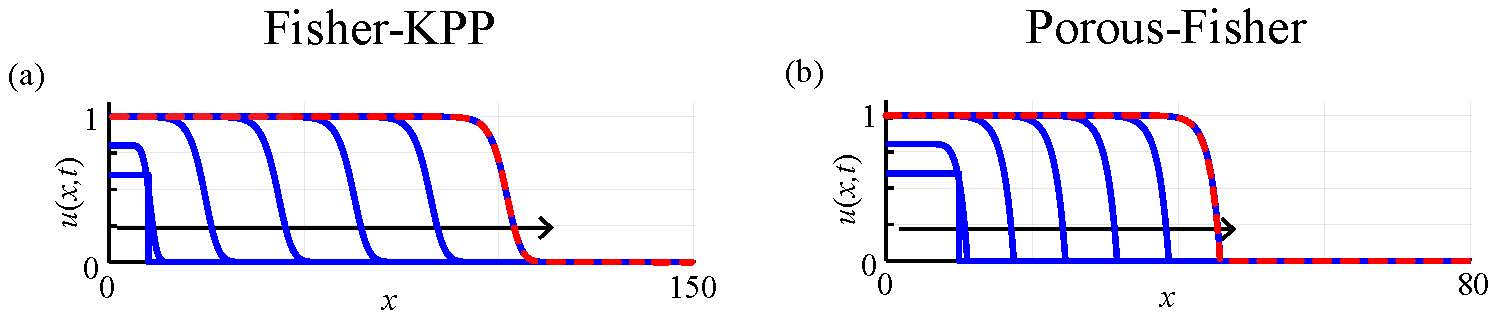}
\renewcommand{\baselinestretch}{1.0}
\caption{Time-dependent solutions of the Fisher-KPP and Porous-Fisher models evolving from initial conditions with compact support, Equation (\ref{eq:FisherICCompactSupport}) with $u_0=0.6$ and $b=10$.  (a) Shows solutions of the Fisher-KPP model computed on $0 \le x \le 150$ with $\delta x = 0.1$ leading to smooth-fronted solutions with $c=1.969520519$. (b) Shows solutions of the Porous-Fisher model computed on $0 \le x \le 80$ with $\delta x=0.01$ leading to sharp-fronted solutions with $c=0.706887186$.  Each plot shows $u(x,t)$ at $t=0,1,10,20,30,40,50$ with the arrow showing the direction of increasing time, and the solution at $u(x,50)$ is superimposed with a red dashed line corresponding to the shifted $U(z)$ profile obtained in the phase plane.}\label{F8}
\end{figure}

Using our time-dependent PDE algorithms we can confirm that long-time shape and speed of the Fisher-KPP and Porous-Fisher solution profiles with the initial condition  (\ref{eq:FisherICCompactSupport}) are independent of the choice of $u_0$ and $b$ (not shown).  For example, typical time-dependent solutions are shown in Figure \ref{F8} where we see that the long-time travelling wave solution of the Fisher-KPP model evolves to the travelling wave solution with the minimum wave speed $c=2$, while the long time travelling wave solution of the Porous-Fisher model also evolves to the travelling wave solution with the minimum wave speed with $c = 1/\sqrt{2}$.

It is useful to know that travelling wave solutions evolving from initial conditions with compact support lead to travelling wave solutions with the minimum wave speed.  For the Fisher-KPP model this is consistent with the observation that  (\ref{eq:FisherICCompactSupport}) can be interpreted as taking $a \to \infty$ in (\ref{eq:FisherInitialCondition}) which leads to $c=2$.  The situation for the Porous-Fisher model is more subtle, as we now explain.  The  results in Figure \ref{F5} for the Porous-Fisher model involve $u(x,0)>0$ for all $x \ge 0$, leading to time-dependent solutions that, despite the apparent sharp front in Figure \ref{F5}(d) strictly speaking all have $u(x,t)>0$ at all $x \ge 0$ and $t \ge 0$.  This is different to the situation in Figure \ref{F8}(b), where we have  $u(x,0)>0$ for $0 < x < b$ and $u(x,0)=0$ for $x > b$.  In this case, the initial condition with compact support leads to a genuine sharp-fronted time-dependent solution that evolves to the sharp-fronted travelling wave (\ref{eq:SharpFrontExact}). Therefore, strictly speaking the results in Figure \ref{F8}(b) are the only initial conditions considered so far where there is a clear front position, and the time-dependent solution in Figure \ref{F8}(b) is the only time-dependent PDE solution where the clear front position persists for $t > 0$.  The case in Figure \ref{F5}(d) where $u(x,0) > 0$ at all $x \ge 0$ is very interesting because we do not expect to have a time-dependent solution with compact support;  we do, however, see the formation of an apparently sharp-fronted travelling wave form at late time, and this travelling wave has the same speed and shape as the minimum speed sharp-front solution (\ref{eq:SharpFrontExact}).  This observation turns out to be a computational artefact arising because of finite precision arithmetic where the far-field values of the initial condition, $u(x,0) \sim u_0 \, \textrm{exp}(-ax)$, are eventually truncated to zero for sufficiently large values of $x$, thereby effectively turning the smooth initial condition into one with compact support.

\section{Fisher-Stefan model}\label{sec:FisherStefan}
The second approach to model travelling wave solutions with a well-defined front is to consider the Fisher-Stefan model~\cite{Du2010,Du2011,ElHachem2019,Elhachem2021,Simpson2020,Elhachem2022,Bui2024} that is obtained by re-formulating the usual Fisher-KPP model as a moving boundary problem:
\begin{equation}
\dfrac{\partial u}{\partial t} = \dfrac{\partial^2 u}{\partial x^2} +  u(1-u), \quad \textrm{on} \quad 0 < x < L(t), \label{eq:FisherStefan}
\end{equation}
with $\partial u / \partial x = 0$ at $x=0$ and $u(L(t),t)=0$.   Here the sharp-front at the leading edge arises because of the homogeneous Dirichlet boundary condition.  In the moving boundary framework, the position of the leading edge is explicitly determined as part of the solution and evolves according to a one-phase Stefan condition
\begin{equation}
\dfrac{\textrm{d}L}{\textrm{d}t} = -\kappa \dfrac{\partial u}{\partial x}, \quad \textrm{at} \quad x=L(t). \label{eq:FisherStefanBC}
\end{equation}
This boundary condition involves specifying a constant, $\kappa$, so that the velocity of the moving front is determined by both the choice of $\kappa$ and $\partial u / \partial x$ at the moving front.  In the heat and mass transfer literature,  the constant $\kappa$ is related to the Stefan number~\cite{Crank1987,Gupta2017}.  All travelling wave solutions of the Fisher--Stefan problem are sharp--fronted, and there is no longer have the possibility of having smooth travelling wave solutions with this modelling framework.

Another point of interest is that the Fisher-KPP and Porous-Fisher models with initial conditions of the form (\ref{eq:FisherInitialCondition}) always lead to \textit{invading} travelling wave solutions $(c > 0$).  In contrast, we will demonstrate that the Fisher-Stefan model supports both invading travelling wave solutions by setting $\kappa > 0$ and receding travelling waves by setting $\kappa < 0$.  For the special case of $\kappa=0$, the Fisher-Stefan model leads to a stationary wave with $c=0$~\cite{ElHachem2019}.  This flexibility in modelling biological invasion and recession is not possible with the Fisher-KPP or Porous-Fisher models.  One interpretation of this difference is that the Fisher-Stefan model allows us to model a wider range of biologically plausible outcomes than either the Fisher-KPP or Porous-Fisher models~\cite{ElHachem2019,Elhachem2022}.

\begin{figure}[htp]
\centering
     \includegraphics[width=1.0\textwidth]{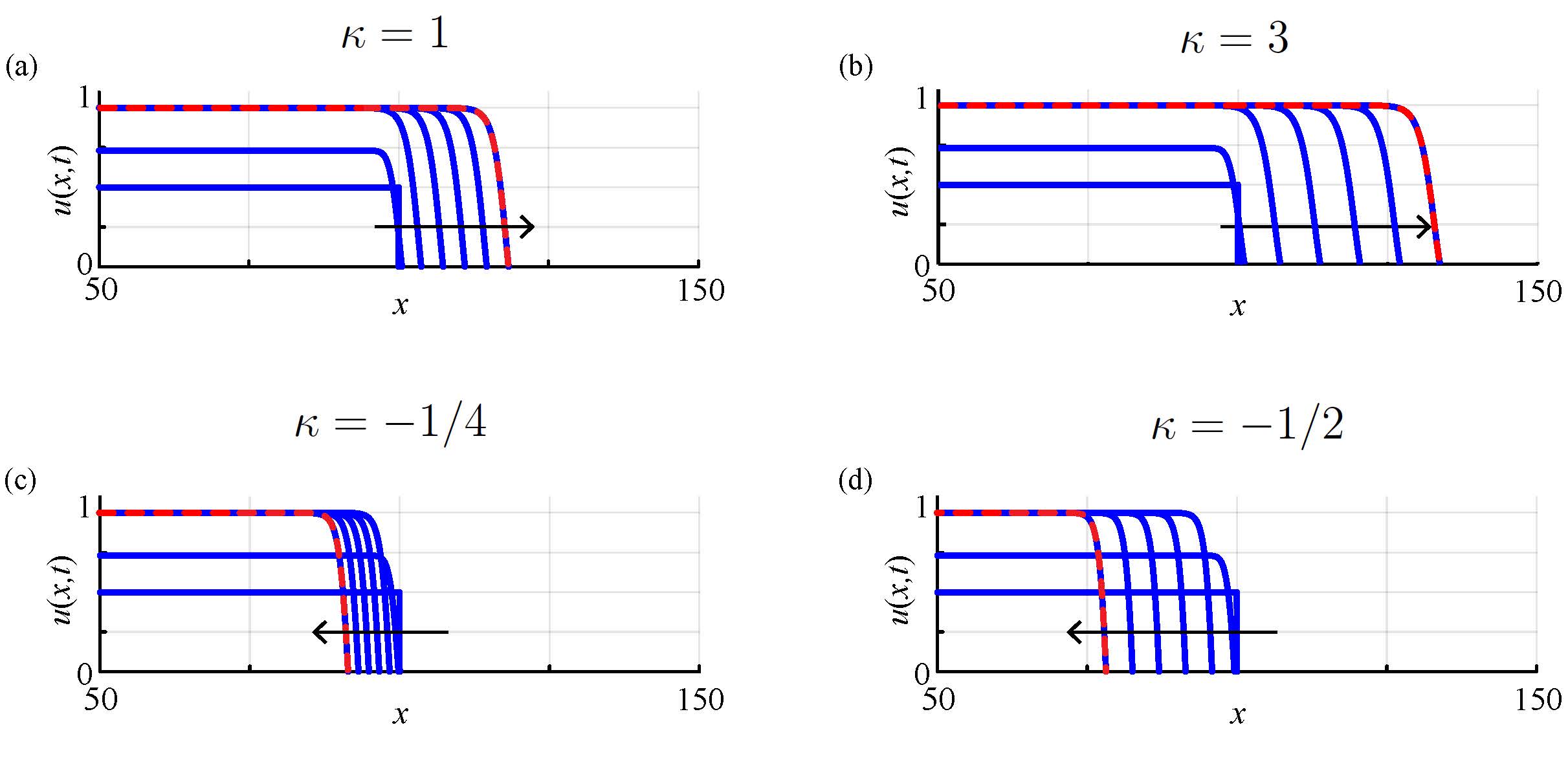}
\renewcommand{\baselinestretch}{1.0}
\caption{Time-dependent solutions of the Fisher-Stefan model illustrate $\kappa$ influences the long-time travelling wave speed $c$.  Each subfigure shows a time-dependent PDE solution with the initial condition given by Equation (\ref{eq:FisherStefanInitialCondition}) with $u_0=0.5$ and $L(0)=100$.  Results show: (a) time-dependent PDE solution for $\kappa=1$ which leads to an invading travelling wave with $c=0.364421881$;  (b) time-dependent PDE solution for $\kappa=3$ which leads to an invading travelling wave with $c=0.665977101$; (c) time-dependent PDE solution for $\kappa=-1/4$ which leads to a receding travelling wave with $c=-0.173023072$; and, (d) time-dependent PDE solution for $\kappa=-1/2$ which leads to a receding travelling wave with $c=-0.442690692$.  Each subfigure shows $u(x,t)$ (solid blue) at $t=0,1,10,20,30,40,50$ with the arrow showing the direction of increasing time. All time-dependent PDE solutions are obtained with $\delta \rho = 2 \times 10^{-4}$, and each time dependent PDE solution at $t=50$ is superimposed with a dashed red line which corresponds to the shifted $U(z)$ profile obtained in the phase plane.} \label{F9}
\end{figure}

It is worth noting that previous research has established various properties of the solution of the Fisher-Stefan model, such as existence and uniqueness~\cite{Du2010,Du2011}, as well as stability considerations~\cite{Bui2024}.  Our approach, as demonstrated for the Fisher-KPP and Porous-Fisher models in Section \ref{sec:FisherSmooth}--\ref{Sec:Compact}, is to focus on using computational methods to explore time-dependent PDE solutions with the view to  studying long-time travelling wave behaviour using a mixture of computational, phase plane and perturbation methods~\cite{ElHachem2019,Elhachem2022,Simpson2020}.  To illustrate our approach, we begin by solving Equation (\ref{eq:FisherStefan}) numerically by specifying initial condition of the form
\begin{equation}
u(x,0)=
    \begin{cases}
        u_0, & \text{if } 0 \le x < L(0), \\
        0,   & \text{if } x = L(0), \label{eq:FisherStefanInitialCondition}
    \end{cases}
\end{equation}
for some choice of $0 < u_0 \le 1$. The first step is to transform the moving boundary problem to a fixed domain by writing $\rho = x / L(t)$, giving
\begin{equation}
\dfrac{\partial u}{\partial t} = \dfrac{1}{L^2}\dfrac{\partial^2 u}{\partial \rho^2}+ \dfrac{1}{L}\dfrac{\textrm{d}L}{\textrm{dt}}\dfrac{\partial u}{\partial \rho} +  u(1-u), \quad \textrm{on} \quad  0 < \rho < 1,
\end{equation}
with $\partial u / \partial \rho = 0$ at $\rho=0$ and $u(1,t)=0$, and
\begin{equation}
\dfrac{\textrm{d}L}{\textrm{d}t} = -\dfrac{\kappa}{L} \dfrac{\partial u}{\partial \rho}, \quad \textrm{at} \quad \rho =1.
\end{equation}
To solve the time-dependent PDE problem we employ a uniform discretisation of $0 < \rho < 1$ with grid spacing $\delta \rho$.   Applying central difference approximations gives
\begin{equation}
\dfrac{\textrm{d}u_i}{\textrm{d}t}=
    \begin{cases}
        \dfrac{u_{i+1}-u_{i}}{(L\delta \rho)^2}+\dfrac{\textrm{d}L}{\textrm{d}t}\left[\dfrac{u_{i+1}-u_{i}}{L \delta \rho}\right]+ u_i (1 - u_i), & \text{if } i=1, \\
        \dfrac{u_{i-1}-2u_{i}+u_{i+1}}{(L\delta \rho)^2}+\dfrac{\textrm{d}L}{\textrm{d}t}\left[\dfrac{u_{i+1}-u_{i-1}}{2L \delta \rho}\right]+  u_i (1 - u_i),    & \text{if } 2 \le i \le N-1, \\
        0, & \text{if } i = N,
    \end{cases}
\end{equation}
which is closed by discretising the moving boundary condition
\begin{align}
\dfrac{\textrm{d}L}{\textrm{d}t} &= -\dfrac{\kappa}{L} \left[ \dfrac{3u_{N} - 4u_{N-1}+ u_{N-2}}{2\delta \rho}\right], \notag \\
&=\dfrac{\kappa}{L} \left[ \dfrac{4u_{N-1}- u_{N-2}}{2\delta \rho}\right], \label{eq:DiscretisedMovingBoundaryCondition}
\end{align}
noting that $u_{N}=0$ because of the homogeneous Dirichlet boundary condition at $\rho = 1$.  Integrating these $N+1$ coupled nonlinear ODEs using DifferentialEquations.jl leads to time-dependent PDE solutions shown in Figure \ref{F9}, each with $u_0 = 0.5$ and $L(0) = 100$.

Simulations in Figure \ref{F9}(a)--(b) for $\kappa > 0$ lead to constant-speed, constant shape invading travelling wave solutions moving in the positive $x$-direction.  The travelling wave speed of these solutions can be estimated numerically like we did in Figure \ref{F2} by tracking the position of some density contour and using regression.  Since the Fisher-Stefan model is a moving boundary problem, there is a more obvious second approach to determine $c$ as we can use our numerical estimates of $u_{N-1}$ and $u_{N-2}$ to evaluate Equation (\ref{eq:DiscretisedMovingBoundaryCondition}) to give a direct estimate of $\textrm{d}L / \textrm{d}t$ without needing to use front tracking and regression. Using this approach at $t=1,2,3,\ldots,50$ indicates that $\textrm{d}L / \textrm{d}t$  settles to a constant speed that agrees very well with the result obtained by front tracking and regression (not shown). In this Section we simply report estimates of $c$ by evaluating (\ref{eq:DiscretisedMovingBoundaryCondition}) using late-time data.  The two invading solutions in Figure \ref{F9}(a)--(b) indicate that $c$ appears to increase with $\kappa$; we will formalise the relationship between these quantities later when we turn to the phase plane.  Interestingly, we have $c \approx 0.36$ for $\kappa = 1$ and $c \approx 0.666$ for $\kappa = 3$, indicating that these two travelling wave speeds are slower than the minimum travelling wave speed for the Fisher-KPP or Porous-Fisher models.  Results in Figure \ref{F9}(c)--(d) for $\kappa < 0$ lead to receding travelling wave solutions with $c < 0$.  Overall, we see that $c$ is an increasing function of $\kappa$ regardless of whether there is an invading or receding travelling wave; again we note that modelling biological recession is simply not possible using the Fisher-KPP or Porous-Fisher models, whereas in the moving boundary framework it is straightforward to model biological recession.

It is very interesting that travelling wave solutions of the Fisher-Stefan PDE model involve the same dynamical system as the Fisher-KPP model, namely
\begin{align}
\dfrac{\textrm{d}U}{\textrm{d}z} &= V, \label{eq:FisherStefanDS1} \\
\dfrac{\textrm{d}V}{\textrm{d}z} &= -cV - U(1-U), \label{eq:FisherStefanDS2}
\end{align}
which we have already established has two equilibrium points $(\bar{U},\bar{V})=(0,0)$ and $(\bar{U},\bar{V})=(1,0)$, where $(\bar{U},\bar{V})=(1,0)$ is a saddle point for all values of $c$.   We have previously established that the origin is a stable spiral for $c < 2$, but we can anticipate from the time-dependent PDE solutions in Figure \ref{F9} that the origin of the phase plane plays no role in these travelling wave solutions since the leading edge of the travelling wave has a negative finite slope, whereas a heteroclinic orbit entering the origin would have zero slope as $U \to 0$.  We will now explore some specific phase planes to provide additional insight.

\begin{figure}[htp]
\centering
     \includegraphics[width=1.0\textwidth]{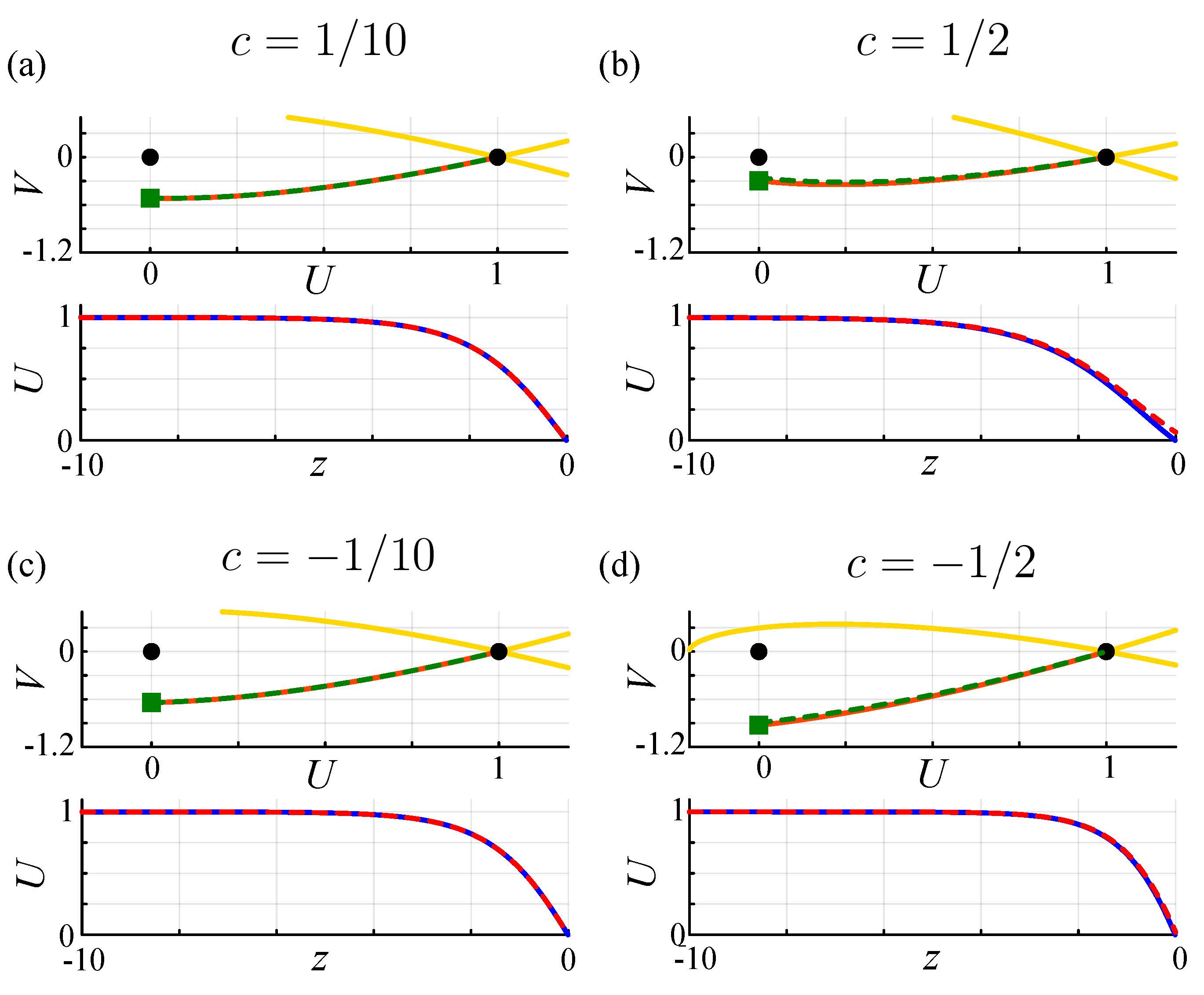}
\renewcommand{\baselinestretch}{1.0}
\caption{Phase planes for (\ref{eq:FisherStefanDS1})--(\ref{eq:FisherStefanDS2}) and approximate perturbation solutions for slowly invading and slowly receding travelling wave solutions for the Fisher-Stefan model with $|c| \ll 1$: (a) $c=0.1$; (b) $c=0.5$; (c) $c=-0.1$; and, (c) $c=-0.5$.  The upper panel in each subfigure shows the phase plane where the equilibria are highlighted (black discs) and the special point $(0, V^\dagger)$ is also highlighted (green square). The red trajectory between $(1,0)$ and $(0,V^\dagger)$ corresponds to the travelling wave solutions and an approximate perturbation solution for $|c| \ll 1$ is superimposed.  Results in the lower panel of each subfigure compare numerical estimates of the shape of the travelling wave (solid blue) with solutions obtained from the approximate perturbation solution (dashed red).} \label{F10}
\end{figure}

Results in Figure \ref{F10} present numerically-generated phase plane trajectories for $c=\pm 0.1$ and $c=\pm 0.5$.  The upper panel in each subfigure shows the phase plane with the two equilibrium points highlighted, and special yellow trajectories entering and leaving the $(1,0)$ equilibrium point along the stable and unstable manifolds.  Each phase plane shows a red trajectory leaving the $(1,0)$ equilibrium point along the unstable manifold into the fourth quadrant of the phase plane until it eventually intersects with the $V$-avis at a special point $(0,V^\dagger)$ which we highlight using a green square.  The trajectory between $(1,0)$ and $(0,V^\dagger)$  corresponds to the travelling wave solution; we have chosen to terminate this trajectory at the point where it intersects with the $V$-axis.  Of course, since $c < 2$, we know from linear stability analysis that the trajectory will eventually spiral into the origin but we do not show this here as this part of the trajectory is not associated with the travelling wave solution from the time-dependent PDE model.

To develop analytical insight into these travelling waves, we re-write (\ref{eq:FisherStefanDS1})--(\ref{eq:FisherStefanDS2}) as a single differential equation that directly describes the shape of the $V(U)$ trajectory in the $(U,V)$ phase plane,
\begin{equation}
V(U)\dfrac{\textrm{d}V(U)}{\textrm{d}U} = -cV(U) - U\left( 1-U\right), \quad V(1)=0.  \label{eq:FisherStefanc01}
\end{equation}
While we are unable to solve (\ref{eq:FisherStefanc01}) to give a general expression for $V(U)$ for arbitrary values of $c$, we can solve for $V(U)$ in the special case $c=0$~\cite{Witelski1994}.  Our approach will be to use this result to construct a perturbation solution of the form
\begin{equation}
V(U) = V_0(U) + c V_1(U) + \mathcal{O}(c^2). \label{eq:FisherStefanc02}
\end{equation}
Substituting (\ref{eq:FisherStefanc02}) into (\ref{eq:FisherStefanc01}) gives
\begin{align}
V_0(U)\dfrac{\textrm{d}V_0(U)}{\textrm{d}U} &= -U(1-U), \quad V_0(1)=0, \label{eq:FisherStefanSmallc}\\
V_0(U)\dfrac{\textrm{d}V_1(U)}{\textrm{d}U} &= -V_0(U) -V_1(U)\dfrac{\textrm{d}V_0(U)}{\textrm{d}U}, \quad V_1(1)=0,
\end{align}
which can be solved to give
\begin{align}
V_0(U)&= \pm \sqrt{\dfrac{2 U^{3}-3 U^{2}+1}{3}} ,\label{eq:FisherStefanSmallcsolution} \\
V_1(U)&= -\frac{\left(\sqrt{2U +1} \left[2U^{2}-3U -2\right]+3\sqrt{3}\right)}{5\sqrt{2 U +1}\, \left(U-1 \right)}. \label{eq:FisherStefanSmallcsolution2}
\end{align}
We choose the negative term in (\ref{eq:FisherStefanSmallcsolution}) since our numerical phase plane trajectories in Figure \ref{F10} involve $V <0$.  Therefore, (\ref{eq:FisherStefanSmallcsolution2}) is relevant for the negative square root in (\ref{eq:FisherStefanSmallcsolution}).  We now have a two--term approximation $V(U) = V_0(U) + c V_1(U)$ describing the trajectory in the phase plane for $|c| \ll 1$.  To illustrate how this approximation performs we superimpose plots of $V_0(U) + c V_1(U)$ in each phase plane in Figure \ref{F10} where we see that this approximate solution is visually indistinguishable from the full numerical solution of the dynamical system for $c = \pm 0.1$.  In contrast, for $c = \pm 0.5$ we see a small difference between the perturbation solution and the full numerically--generated trajectory, as we might expect.

To explore the relationship between $\kappa$ and $c$, we evaluate our two-term perturbation expression at $U=0$ to derive an expression for $V(0) = V^\dagger$.  Substituting this expression into the moving boundary condition
\begin{equation}
\dfrac{\textrm{d}L}{\textrm{d}t}= -\kappa V^{\dagger},
\end{equation}
gives a relationship between $\kappa$ and $c$, noting that $\textrm{d}L/\textrm{d}t \to c$ as $t \to \infty$.  Rearranging this expression gives
\begin{equation}
\kappa =  \sqrt{3}c+ \left(\frac{9 \sqrt{3} - 6}{5}\right) c^2 + \mathcal{O}(c^3),
\end{equation}
confirming that  $c$ increases with $\kappa$ for both slowly invading and receding travelling wave solutions with $|c| \ll 1$.

It is interesting to note that we have made analytic progress for slow travelling wave solutions by developing approximate expressions for certain trajectories in the phase plane rather than working directly to obtain approximate expressions for $U(z)$ like we were able to do for fast travelling wave solutions of the Fisher-KPP and Porous-Fisher models when $c \gg  1$.  Our approach here of working in the phase plane is advantageous since we are able to use these perturbation solutions to derive an expression relating $c$ to $\kappa$ through the moving boundary condition.  To compare the shape of the travelling wave solution $U(z)$ with our time-dependent PDE solutions, we take our numerical estimates of $U(V)$ and recall that $\textrm{d}U/\textrm{d}z = V$ by definition.  Integrating this ODE numerically, again using the trapezoid rule, provides us with estimates of $U(z)$.  This approach can be used to determine $U(z)$ either using the numerically generated  phase plane trajectory for any value of $c$ or the two-term perturbation approximation for $|c| \ll 1$.  Using this approach we computed $U(z)$, and compared the two profiles in the lower panel of each subfigure in Figure \ref{F10}.  Here we see that the perturbation approach leads to very accurate profiles of $U(z)$ for $c=\pm 0.1$.  We also note that the late-time  PDE solutions in Figure \ref{F8} are superimposed with red dashed curves that are indistinguishable from the late-time PDE solution.  These red dashed curves are obtained by generating the phase plane for the approximate value of $c$ and using the $V(U)$ trajectory in the phase plane to estimate the $U(z)$ curve in the same way as described here.

Numerical simulations of time-dependent PDE solutions in Figure \ref{F10}(c)--(d) show PDE  solutions with $\kappa < 0$ that lead to receding travelling wave solutions with $c < 0$.   The shape of the $V(U)$ trajectories in the phase plane are given by Equation (\ref{eq:FisherStefanc01}), which has the property that $\textrm{d}V/\textrm{d}U = -c + U(1-U)/V \sim -c$ as $c \to -\infty$, suggesting that the relevant phase plane trajectories approach a straight line $V = c(U-1)$ as $c \to -\infty$.  For this limiting straight line trajectory we have $V(0) = V^\dagger = -c$, indicating that $\kappa \to -1^+$ as $c \to -\infty$.  Note that for $\kappa < -1$, instead of evolving to a travelling wave profile, the time-dependent solution of (\ref{eq:FisherStefan})--(\ref{eq:FisherStefanBC}) may undergo a form of finite-time blow-up~\cite{McCue2022}, which is not the subject of this review.

To explore the shape of the travelling wave solution as $c \to -\infty$ we consider
\begin{equation}
\dfrac{1}{c}\dfrac{\textrm{d}^2U}{\textrm{d}z^2}+ \dfrac{\textrm{d}U}{\textrm{d}z} + \dfrac{1}{c}U(1-U)=0,
\end{equation}
which is singular as $c \to -\infty$ and motivates us to consider a matched asymptotic expansion~\cite{Murray1984} treating $c^{-1}$ as a small parameter.  The boundary conditions for this problem are $U(0)=0$ and $U(z)=1$ as $z \to -\infty$.  Setting $c^{-1}=0$ and solving the resulting ODE gives the outer solution
\begin{equation}
U(z)=1,
\end{equation}
to match the far--field boundary condition.  We construct the inner solution near $z=0$ by re-scaling the independent variable $\eta = cz$, giving
\begin{equation}
\dfrac{\textrm{d}^2U}{\textrm{d}\eta^2}+ \dfrac{\textrm{d}U}{\textrm{d}\eta} + \dfrac{1}{c^2}U(1-U)=0, \quad \textrm{on} \quad -\infty < \eta < 0.
\end{equation}
Treating $c^{-1}$ as a small parameter, we seek a perturbation solution of the form $U(\eta) = U_0(\eta)+c^{-2} U_1(\eta) + \mathcal{O}(c^{-4})$, giving
\begin{align}
& \dfrac{\textrm{d}^2U_0(\eta)}{\textrm{d}\eta^2} +\dfrac{\textrm{d}U_0(\eta)}{\textrm{d}\eta} =0, \quad U_0(0)=0, \quad U_0(\eta) =1 \quad \textrm{as} \quad \eta \to \infty,\\
& \dfrac{\textrm{d}^2U_1(\eta)}{\textrm{d}\eta^2} +\dfrac{\textrm{d}U_1(\eta)}{\textrm{d}\eta} + U_0(\eta)(1-U_0(\eta)) =0, \quad U_1(0)=0, \quad U_1(\eta) =0 \quad \textrm{as} \quad \eta \to \infty.
\end{align}
The solution of these two boundary value problems, written in terms of the $z$ variable is
\begin{align}
U_0(z) &= 1 - \textrm{e}^{-2cz},\\
U_1(z) &= \dfrac{1}{2}\left(\textrm{e}^{-2cz} + (2zc - 1)\textrm{e}^{-cz}\right),
\end{align}
gives us a two-term perturbation solution that directly approximates the shape of the travelling wave profile.

\begin{figure}[htp]
\centering
     \includegraphics[width=1.0\textwidth]{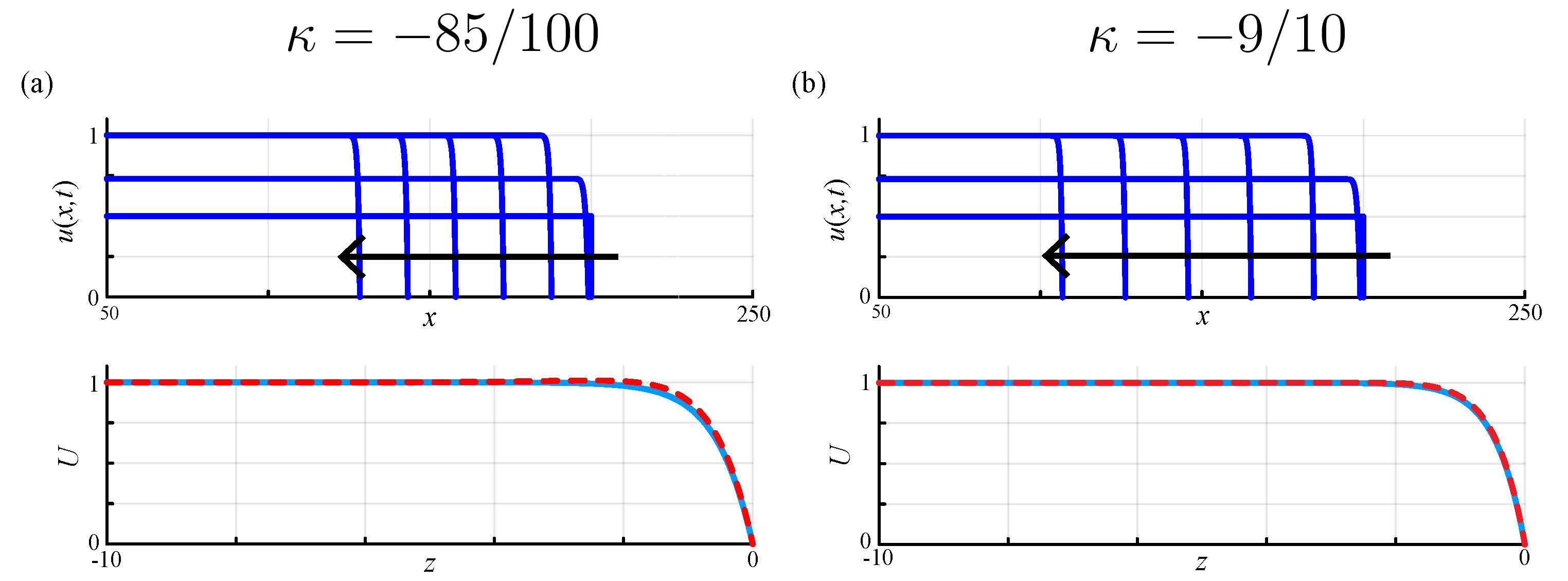}
\renewcommand{\baselinestretch}{1.0}
\caption{Fast receding travelling wave solutions for (a) $\kappa=-0.85$; and (b) $\kappa=-0.90$.  Results in the upper panel show time-dependent PDE solutions evolving from $u_0=0.5$ and $L(0)=200$.  Profiles are shown at $t=0,1,10,20,30,40,50$ with the arrow showing the direction of increasing time.  These numerical results indicate that we have $c=-1.482746707$ for $\kappa = -0.85$ and  $c=-1.948448670$ for $\kappa = -0.90$.  Results in the lower-panel compare the shape of the travelling wave profiles estimated from late-time PDE solutions (solid blue) and the two-term perturbation solution (dashed red).  All time-dependent PDE solutions are obtained with $\delta \rho = 2 \times 10^{-4}$} \label{F11}
\end{figure}

Recalling that $c \to -\infty$ as $\kappa \to -1^+$, we generate time-dependent PDE solutions, shown in Figure \ref{F11}, for $\kappa = -0.85$ and $\kappa = -0.90$.  In both cases we see that the initial profile forms a receding travelling wave with a steep boundary layer forming near $x = L(t)$.  In each figure we plot a late-time solution, focussing in on the shape of the leading edge over the interval $-10 \le z \le 0$ and compare the shape of the receding travelling wave generated by the time-dependent PDE solver with our two-term perturbation solution.  As expected, we see some small discrepancy between the profiles when $\kappa = -0.85$, but the numerical and perturbation results are visually indistinguishable, at this scale, for $\kappa = -0.90$.  The accuracy of the two-term matched asymptotic perturbation solution in Figure \ref{F11}(b) is remarkable given that this approximation is valid in the limit $c \to -\infty$ and yet we find it provides a reasonable approximation for just $c \approx -1.95$.

In addition to providing an understanding of the shape of fast receding travelling wave solutions, our perturbation solutions also provide insight into the relationship between $c$ and $\kappa$ as $c \to -\infty$.    For all travelling wave solutions we have $\kappa = -c / V(0)$, and as $c \to -\infty$ we can estimate $V(0)$ using our perturbation solutions by evaluating $\kappa = -c / \textrm{d}U/\textrm{d}z$ at $z=0$ by differentiating our expressions for $U_0(z)$ and $U_1(z)$ with respect to $z$ and evaluating the resulting expressions at $z=0$, giving
\begin{equation}
\kappa = -1 + \dfrac{1}{2c^2} + \mathcal{O}\left(\dfrac{1}{c^4}\right),
\end{equation}
which shows precisely how $\kappa$ relates to $c$ as $c \to -\infty$. These results for $c \to -\infty$ correspond to $\kappa \to -1^+$.  Time-dependent PDE solutions can be generated for $\kappa \le -1$, but in this case we see the formation of receding fronts that appear to accelerate in the negative $x$-direction rather than settle to a constant speed travelling wave solution~\cite{McCue2022}.

Before concluding our discussion about the Fisher-Stefan model it is insightful to  note that using both the time-dependent PDE model and phase plane tools provides us with opportunities to independently check the accuracy of our numerical algorithms.  For example, generating the time-dependent PDE solutions in Figure \ref{F9} involves specifying $\kappa$ as an input and observing $c$ as an output.  This process involves several different approximations since we must discretise the governing PDE, solve the resulting system of semi-discrete ODEs for a sufficiently long period of time before using the discretised boundary condition to provide estimate $c$.  For example, time-dependent solutions in Figure \ref{F9}(a) involve specifying $\kappa = 1$, and our estimate of  $c=0.364421881$ is impacted by several choices we have made in implementing our numerical algorithm, such as truncation error.  One way of testing the accuracy of our approximation is to repeat the calculation by re-solving the time-dependent PDE model on a finer mesh and examine how our estimate of $c$ depends upon the choice of $\delta \rho$.  A different approach is to use the phase plane where we specify $c$ as an input, generate the trajectory leaving $(1,0)$ along the unstable manifold and to use this trajectory to estimate $V^\dagger$, from which we can apply the moving boundary condition, $c = -\kappa V^\dagger$ to give $\kappa$ as an output of the phase plane calculations.  To demonstrate this approach we generated the phase plane with $c=0.364421881$ as an input, solved for the appropriate trajectory to give $V^\dagger = -0.365356934$, and then applied the moving boundary condition to find that $\kappa = 0.997440708$, which is within 0.26\% of the expected result, $\kappa = 1$.  This inter-code comparison is convenient since it gives us confidence that both the time-dependent PDE algorithm and the numerical phase plane algorithm are giving consistent results.  In addition, if we repeat this inter-code comparison by re-solving the time-dependent PDE model with more mesh points we obtain an improved estimate of $c$ which, when used as an input into the phase plane, leads to an improved estimate of $\kappa$ (results not shown).

Another point of interest about the Fisher-Stefan model is that our results in Figure \ref{F9} all lead to long-time travelling wave solutions.  There is another possibility that we have not focused on in this review, which is motivated by the observation that the boundary condition $u(L(t),t) = 0$ is associated with a diffusive loss of material at $x=L(t)$, where the outward flux is $\mathcal{J} = -\partial u / \partial x$.  This loss of material can lead to the population going extinct and $u \to 0$ as $t \to \infty$~\cite{Du2010,ElHachem2019,Simpson2020}.  These two different outcomes have given rise to what has been called the \textit{spreading-vanishing dichotomy} for the Fisher-Stefan model~\cite{Du2010,ElHachem2019,Simpson2020}.  The eventual growth or extinction of the initial population is governed by the diffusive loss at $u=L(t)$ and the increase in mass driven by the logistic source term, and the balance of these two processes is determined by the initial length of the population, $L(0)$.  Simple conservation arguments show that $L(0) > \pi /2$ lead to spreading and long-time travelling wave behaviour, whereas $L(0) < \pi / 2$ leads to eventual extinction of the population.  Since this review focuses on travelling wave solutions,  we have intentionally focused on time-dependent PDE solutions with $L(0) > \pi /2$; however, our time-dependent PDE codes available on \href{https://github.com/ProfMJSimpson/PDEInvasion}{GitHub} can be used to explore different initial conditions leading to extinction simply by changing $L(0)$.  Details of the analysis explaining how to arrive at the critical value of $L(0) = \pi /2$ is provided in our previous work~\cite{ElHachem2019,Simpson2020}.

\section{Discussion}
In this review we have presented open access computational tools written in Julia for three reaction-diffusion models of biological invasion, namely the Fisher-KPP, Porous-Fisher and Fisher-Stefan models.  The first of these, the Fisher-KPP model, is very well-studied in the literature, while the latter two can be motivated in part to address certain limitations of the Fisher-KPP model, including the preference for well-defined fronts and the possibility of invading of receding populations. We anticipate that these computational tools could be used to support senior undergraduate and graduate-level teaching and learning about mathematical modelling of biological invasion.  Moreover, we anticipate that the computational tools will be straightforward to adapt to other reaction-diffusion models, as required.  For each PDE model we generate time-dependent solutions, paying particular attention to relate late-time PDE solutions to properties of the travelling wave phase plane.  Using perturbation methods we provide mathematical insight into the shape of various travelling wave solutions in certain  asymptotic limits.  Using a range of computation and visualisation, we relate these approximate solutions to numerical solutions of the full PDE problems. Finally, in Appendix A, we include equivalent details for an additional model, the Porous-Fisher-Stefan model, that combined features of the Porous-Fisher and Fisher-Stefan models.

This review deliberately focuses on three important types of reaction-diffusion models of biological invasion.  It is obvious that there are many opportunities for extending the material in this review.  For example, we do not discuss any source terms other than logistic growth; however, it is relatively straightforward to implement different forms of sigmoid source terms in the time-dependent PDE algorithms, such as the broad family of generalised logistic source terms reviewed by Tsoularis and Wallace~\cite{Tsoularis2002}.  These modified algorithms can be used to explore how the resulting travelling wave solutions and their interpretation in the phase plane.  Another clear restriction in this review is that we only consider nonlinear degenerate diffusion of the form $\mathcal{J} = - u \, \partial u / \partial x$, whereas others have considered a more general terms such as $\mathcal{J} = - u^m \, \partial u / \partial x$ for some exponent $m > 0$~\cite{Sherratt1990,Jin2016,McCue2019}.  Both of these generalisations can be explored by adapting the open source computational tools provided as part of this review.

A more mathematically challenging but biologically realistic extension of the work covered in this review is to relax the constraint of focusing on scalar reaction--diffusion models.  Working with coupled problems, sometimes called \textit{multispecies} models, brings new opportunities for developing more realistic mathematical models as real applications of biological invasions typically involve interactions of multiple subpopulations.  Multispecies models have been formulated and analysed from the point of view of considering general interactions between different subpopulations~\cite{Painter2003} as well as more specific models of biological processes such as trophoblast invasion during pregnancy~\cite{Landman1998}, malignant invasion into surrounding tissues~\cite{Gatenby1996,Browning2019} and many more.  Such multi-species models can be formulated either as classical PDE problems that directly generalise the Fisher-KPP model~\cite{Haridas2017a} or as multi-species analogues of moving boundary problems similar to the Fisher-Stefan model~\cite{Elhachem2020}.  While extending the computational tools provide here for solving time-dependent PDE models is relatively straightforward for multispecies models, our ability to extend the phase plane tools becomes limited.  For example, travelling wave solutions arising in a two--species reaction--diffusion model of biological invasion typically leads to a four-dimensional phase space problem which is far more difficult to analyse as a dynamical system and more challenging to visualise than types of simpler phase planes explored here.

\noindent
\textbf{Acknowledgements}  Key algorithms used to generate results are available on \texttt{\href{https://github.com/ProfMJSimpson/PDEInvasion}{GitHub}}.  This work is supported by the Australian Research Council (DP230100025).

\newpage
\section{Appendix: Porous--Fisher--Stefan model}
As we point out in the main document, two common approaches for developing simple reaction-diffusion  models that lead to sharp-fronted travelling waves are to either introduce a degenerate nonlinear diffusion mechanism into the classical Fisher-KPP model or to re-formulate the Fisher-KPP model as a moving boundary problem.  Interestingly, it is also possible to combine both of these approaches by reformulating the Porous-Fisher model as a moving boundary problem,
\begin{equation}
\dfrac{\partial u}{\partial t} = \dfrac{\partial}{\partial x}\left[u\dfrac{\partial u}{\partial x}\right] + u(1-u), \quad \textrm{on} \quad 0 < x < L(t), \label{eq:PFSPDE}
\end{equation}
with $\partial u / \partial x = 0$ at $x=0$ and $u(L(t),t)=0$. As for the Fisher-Stefan model we have a moving boundary problem where $L(t)$ is determined as part of the solution by specifying~\cite{Fadai2020}
\begin{equation}
\dfrac{\textrm{d}L}{\textrm{d}t} = -\kappa u \dfrac{\partial u}{\partial x} \quad \textrm{at} \quad x=L(t). \label{eq:PFSBoundaryCondition}
\end{equation}
This model has been referred to as the Porous-Fisher-Stefan model~\cite{Fadai2020} Similar to the Fisher-Stefan model, here the moving boundary condition involves specifying a constant $\kappa$ that determines the velocity of the leading edge.   The Porous-Fisher-Stefan model is closely related to both the Porous-Fisher and Fisher-Stefan models; accordingly our analysis of travelling wave solutions of the Porous-Fisher-Stefan model will borrow ideas previously employed for the Porous-Fisher and Fisher-Stefan models.  As we will show later, as the velocity of the moving boundary in the Porous-Fisher-Stefan model is finite, (\ref{eq:PFSBoundaryCondition}) implies that the gradient of the density profile, $\partial u / \partial x$, must be infinite at the leading edge as we have $u(L(t),t)=0$.

\begin{figure}[htp]
\centering
     \includegraphics[width=1.0\textwidth]{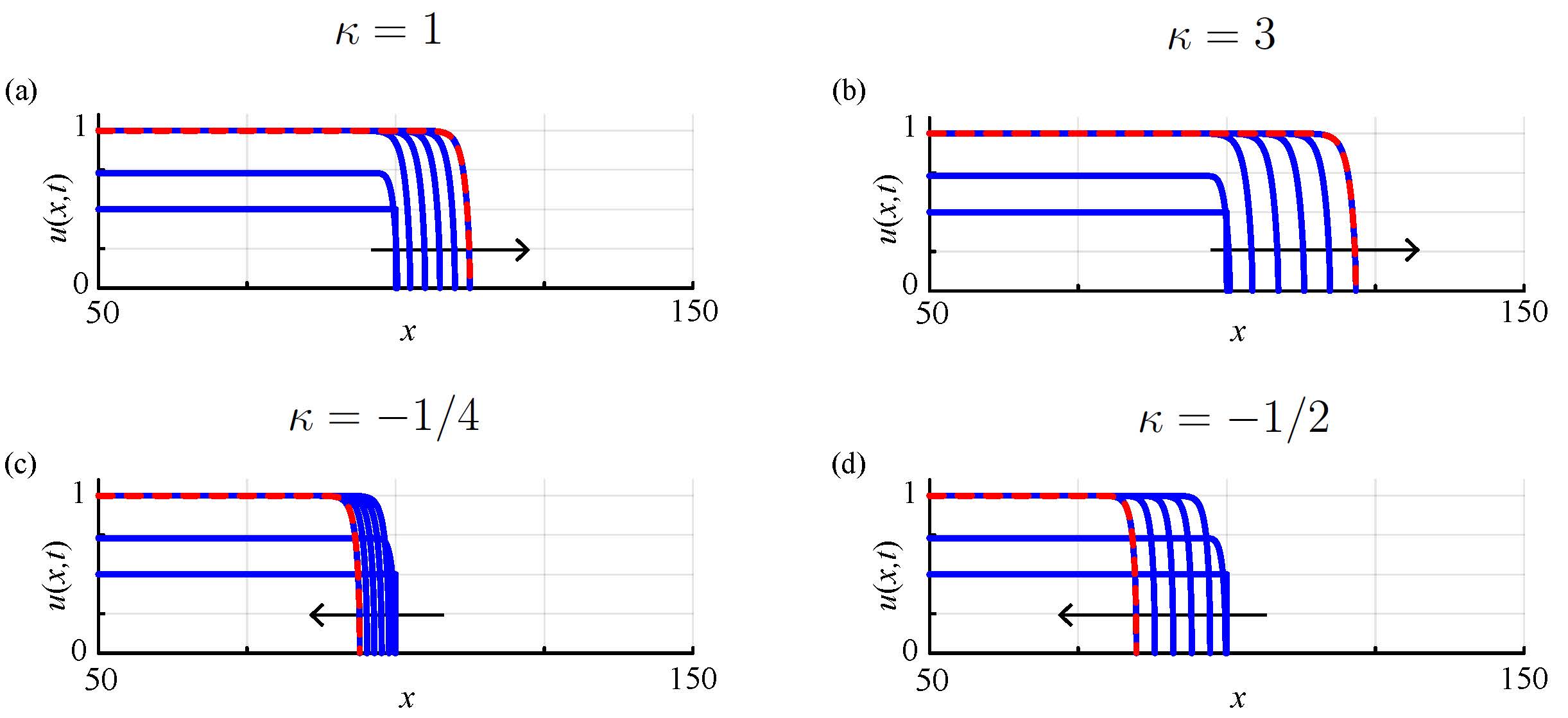}
\renewcommand{\baselinestretch}{1.0}
\caption{Time-dependent solutions of the Porous-Fisher-Stefan model illustrate $\kappa$ influences the long-time travelling wave speed $c$.  Each subfigure shows a time-dependent PDE solution with the initial condition given by Equation (\ref{eq:FisherStefanInitialCondition}) with $u_0=0.5$ and $L(0)=100$.  Results show: (a) time-dependent PDE solution for $\kappa=1$ which leads to an invading travelling wave with $c=0.251165418$;  (b) time-dependent PDE solution for $\kappa=3$ which leads to an invading travelling wave with $c=0.435671955$; (c) time-dependent PDE solution for $\kappa=-1/4$ which leads to a receding travelling wave with $c=-0.122673575$; and, (d) time-dependent PDE solution for $\kappa=-1/2$ which leads to a receding travelling wave with $c=-0.310757910$.  Each subfigure shows $u(x,t)$ (solid blue) at $t=0,1,10,20,30,40,50$ with the arrow showing the direction of increasing time. All time-dependent PDE solutions are obtained with $\delta \rho = 2 \times 10^{-4}$, and each time dependent PDE solution at $t=50$ is superimposed with a dashed red line which corresponds to the shifted $U(z)$ profile obtained in the phase plane.} \label{F12}
\end{figure}

To explore time-dependent solutions of the Porous-Fisher-Stefan model, we introduce a transformed variable $\phi  = u^2$, for $\phi \ge 0$, which gives
\begin{equation}
\dfrac{\partial \phi}{\partial t} = \sqrt{\phi}\dfrac{\partial^2 \phi}{\partial x^2} + 2  \phi \left(1 - \sqrt{\phi} \right), \quad \textrm{on} \quad 0 < x < L(t),
\end{equation}
with $\partial \phi  / \partial x = 0$ at $x=0$ and $\phi(L(t),t)=0$.  Re-writing the time-dependent PDE in terms of $\phi$ may, at first, seem unhelpful since the PDE is re-cast in an unfamiliar form that cannot be written as a standard conservation PDE.  The benefit of working in the $\phi$ variable becomes clear when we consider the moving boundary condition
\begin{equation}
\dfrac{\textrm{d}L}{\textrm{d}t} = -\dfrac{\kappa}{2} \dfrac{\partial \phi}{\partial x} \quad \textrm{at} \quad x=L(t),
\end{equation}
which avoids the $0 \times  \infty$ indeterminate form in (\ref{eq:PFSBoundaryCondition}) for the original $u$ variable. To solve the moving boundary problem for $\phi(x,t)$ numerically, we first transform to the fixed boundary by writing $\rho = x / L(t)$ which gives
\begin{equation}
\dfrac{\partial \phi}{\partial t} = \dfrac{\sqrt{\phi}}{L^2}\dfrac{\partial^2 \phi}{\partial \rho^2} + \dfrac{1}{L}\dfrac{\textrm{d}L}{\textrm{d}t}\dfrac{\partial \phi}{\partial \rho} + 2  \phi \left(1 - \sqrt{\phi} \right),  \quad \textrm{on} \quad 0 < \rho < 1,
\end{equation}
with $\partial \phi  / \partial \rho = 0$ at $\rho=0$ and $\phi(1,t)=0$.  In the fixed-domain coordinates we have
\begin{equation}
\dfrac{\textrm{d}L}{\textrm{d}t} = -\dfrac{\kappa}{2L} \dfrac{\partial \phi}{\partial \rho} \quad \textrm{at} \quad \rho=1.
\end{equation}
To solve the time-dependent PDE problem we consider a uniform discretisation of $0 < \rho < 1$ with grid spacing $\delta \rho$.  Applying central difference approximations gives
\begin{equation}
\dfrac{\textrm{d}\phi_i}{\textrm{d}t}=
    \begin{cases}
\dfrac{\sqrt{\phi_i}(\phi_{i+1}-\phi_{i})}{(L\delta \rho)^2}+\dfrac{\textrm{d}L}{\textrm{d}t}\left[\dfrac{\phi_{i+1}-\phi_{i}}{L \delta \rho}\right]+ 2 \phi_i \left(1 - \sqrt{\phi_i} \right),  & \text{if } i=1, \\
        \dfrac{ \sqrt{\phi_i}(\phi_{i-1}-2\phi_{i}+\phi_{i+1})}{(L\delta \rho)^2}+\dfrac{\textrm{d}L}{\textrm{d}t}\left[\dfrac{\phi_{i+1}-\phi_{i-1}}{2L \delta \rho}\right]+ 2 \phi_i \left(1 - \sqrt{\phi_i} \right),    & \text{if } 2 \le i \le N-1, \\
        0, & \text{if } i = N,
    \end{cases}
\end{equation}
which is closed by specifying
\begin{align}
\dfrac{\textrm{d}L}{\textrm{d}t} &= -\dfrac{\kappa}{2L} \left[ \dfrac{3\phi_{N} - 4\phi_{N-1}+ \phi_{N-2}}{2\delta \rho}\right], \notag \\
&=\dfrac{\kappa}{2L} \left[ \dfrac{4\phi_{N-1}-u_{N-2}}{2\delta \rho}\right],
\end{align}
noting that $\phi_{N}=0$ by definition.  This system of $N+1$ coupled nonlinear ODEs can be integrated through time using DifferentialEquations.jl to give estimates of $\phi(\rho_i,t)$ and $L(t)$.  Recalling that $u = +\sqrt{\phi}$,  we can plot the solution in terms of $u(x,t)$.  Note that the appropriate boundary condition on the moving boundary is to enforce $u(L(t),t) = 0$.  Computationally, however, it necessary to approximate this as  $u(L(t),t) = \mathcal{U}$, with  $\mathcal{U} \ll 1$ otherwise the discretised system of ODEs does not evolve with time.  All numerical results in Figure \ref{F12} are obtained by setting $\phi(L(t),t) = 1 \times 10^{-8}$.  These time-dependent PDE results are reminiscent of those in Figure \ref{F9} for the Fisher-Stefan problem as we see a series of sharp-fronted moving fronts that, in the long-time limit, approach a constant-speed and constant-shape travelling wave.  Moreover, we see that setting $\kappa > 0$ leads to invading travelling waves with $c > 0$, whereas setting $\kappa < 0$ we obtain receding travelling wave with $c < 0$.

To study the travelling wave solutions of the Porous-Fisher-Stefan model we turn to the  phase plane.  Working in the travelling wave coordinate $z=x-ct$,  we note that travelling wave solutions are invariant with respect to a translation in $z$ so we are free to choose $z=0$ to represent the location of the moving boundary which gives
\begin{equation}
\sqrt{\phi} \dfrac{\textrm{d}^2\phi}{\textrm{d}z^2}+ c \dfrac{\textrm{d}\phi}{\textrm{d}z} + 2\phi\left(1-\sqrt{\phi}\right)=0, \quad \textrm{on} \quad -\infty < z < 0, \label{eq:PFSTravellingWave}
\end{equation}
with boundary conditions
\begin{equation}
\lim_{z \to -\infty}\phi(z) = 1, \quad \phi(0) = 0, \quad \textrm{and} \quad \lim_{z \to 0^-}\dfrac{\textrm{d}\phi}{\textrm{d}z} = -\dfrac{2c}{\kappa}. \label{eq:PFSTravellingWaveBC}
\end{equation}
We explore solutions to this boundary value problem by considering the $(\phi,\psi)$ phase plane
\begin{align}
\dfrac{\textrm{d}\phi}{\textrm{d}z} &= \psi, \label{eq:PhasePlane PFS1}\\
\dfrac{\textrm{d}\psi}{\textrm{d}z} &= -\dfrac{c\psi}{\sqrt{\phi}} - 2\sqrt{\phi}(1-\sqrt{\phi}), \label{eq:PhasePlane PFS2}
\end{align}
which is equivalent to the dynamical system for travelling wave solutions of the Porous-Fisher model written in terms of the $(\phi,\psi)$ variables instead of the usual $(U,V)$ variables like we had in Section \ref{sec:PorousFisherSmooth}.  We are unable to find exact solutions of (\ref{eq:PhasePlane PFS1})-(\ref{eq:PhasePlane PFS2}) for arbitrary values of $c$. Instead, inspired by our approach for studying travelling waves of the Fisher-Stefan model, we re-write (\ref{eq:PhasePlane PFS1})-(\ref{eq:PhasePlane PFS2}) as an ODE that describes the trajectory shape $\psi(\phi)$, in the $(\phi,\psi)$ phase plane giving
\begin{equation}
\dfrac{\textrm{d}\psi(\phi)}{\textrm{d}\phi} = -\dfrac{c}{\sqrt{\phi}}-\dfrac{2\sqrt{\phi}\left(1-\sqrt{\phi}\right)}{\psi(\phi)}, \quad \psi(1) = 0, \quad \lim_{\phi \to 0^+}\psi(\phi) = -\dfrac{2c}{\kappa}. \label{eq:PFSPhasePlaneShape}
\end{equation}
As noted in Section \ref{sec:PorousFisherSmooth}, travelling wave solutions of the Porous-Fisher model only occur for $c \ge 1 / \sqrt{2}$ and our numerical exploration of time-dependent PDE solutions for the Porous-Fisher-Stefan model in Figure \ref{F12} suggests that we have travelling wave solutions with $c < 1/\sqrt{2}$.  Guided by our experience with the Fisher-Stefan model in Section \ref{sec:FisherStefan},  we note that it is possible to find an exact solution of Equation (\ref{eq:PFSPhasePlaneShape}) for the special case of $c=0$.  These two observations motivate us to consider (\ref{eq:PFSPhasePlaneShape}) in two asymptotic limits: $c \to 0$ and $c \to 1/\sqrt{2}^{\, -}$.

\begin{figure}[htp]
\centering
     \includegraphics[width=1.0\textwidth]{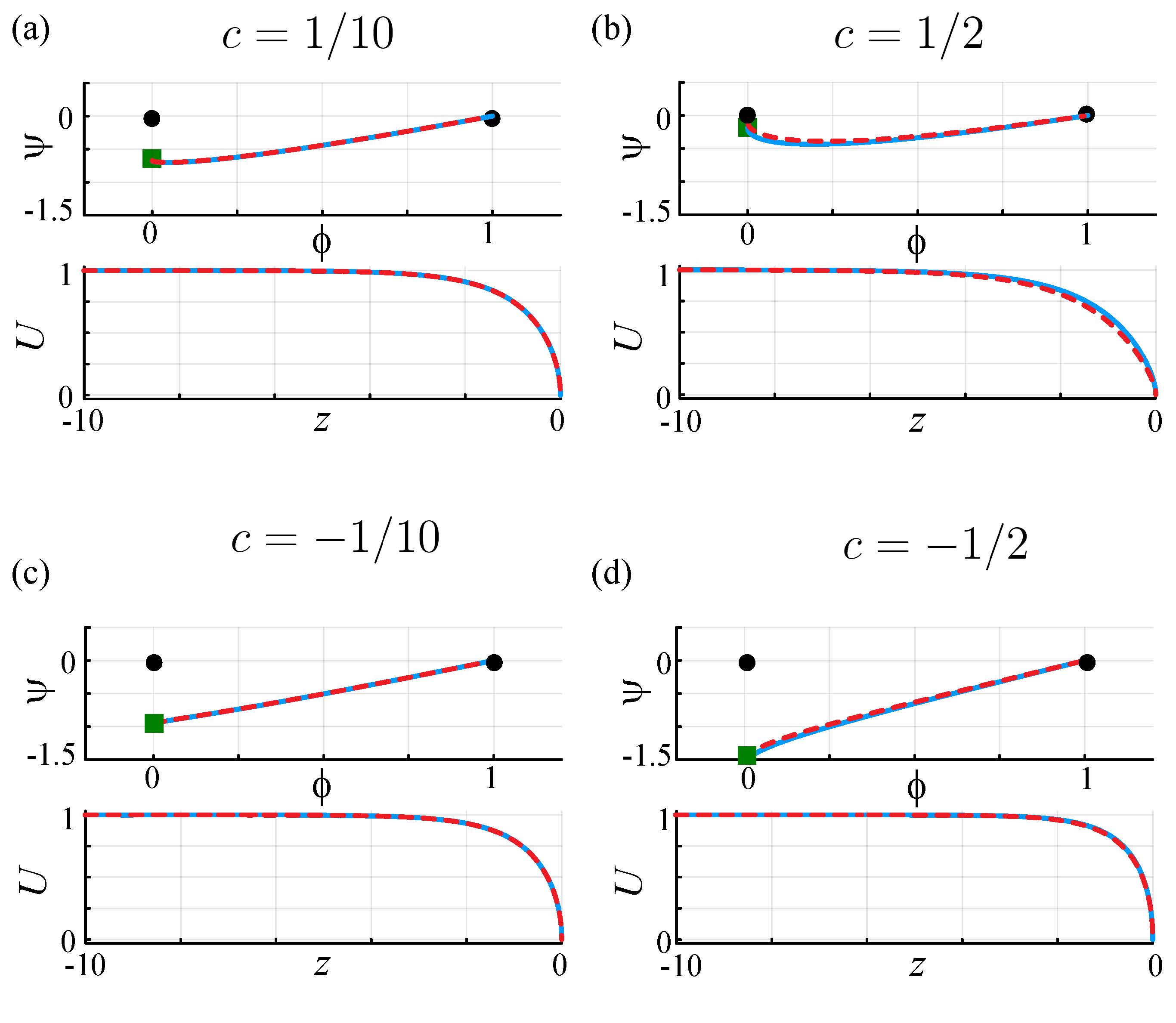}
\renewcommand{\baselinestretch}{1.0}
\caption{Phase planes for (\ref{eq:PhasePlane PFS1})--(\ref{eq:PhasePlane PFS2}) and approximate perturbation solutions for slowly invading and slowly receding travelling wave solutions for the Porous-Fisher-Stefan model with $|c| \ll 1$: (a) $c=0.1$; (b) $c=0.5$; (c) $c=-0.1$; and, (c) $c=-0.5$.  The upper panel in each subfigure shows the phase plane where the equilibria are highlighted (black discs) and the special point $(0, \psi^\dagger)$ is also highlighted (green square). The red trajectory between $(1,0)$ and $(0,\psi^\dagger)$ corresponds to the travelling wave solutions and an approximate perturbation solution for $|c| \ll 1$ is superimposed.  Results in the lower panel of each subfigure compare numerical estimates of the shape of the travelling wave (solid blue) with solutions obtained from the approximate perturbation solution (dashed red).} \label{F13}
\end{figure}

To explore solutions of Equation (\ref{eq:PFSPhasePlaneShape}) for $|c| \ll 1$ we expand $\psi(\phi)$ in a regular perturbation expansion
\begin{equation}
\psi(\phi) = \psi_0(\phi) + c \psi_1(\phi) + \mathcal{O}(c^2),
\end{equation}
which, when substituted into  (\ref{eq:PFSPhasePlaneShape}) gives,
\begin{align}
\psi_0(\phi)\dfrac{\textrm{d}\psi_0(\phi)}{\textrm{d}\phi} &= -2\sqrt{\phi}\left(1-\sqrt{\phi}\right), \quad \psi_0(1)=0, \label{eq:PorousFisherStefanSmallc1}\\
\psi_0(\phi)\dfrac{\textrm{d}\psi_1(\phi)}{\textrm{d}\psi} &=  -\psi_1(\phi)\dfrac{\textrm{d}\psi_0(\phi)}{\textrm{d}\psi} - \dfrac{\psi_0(\phi)}{\sqrt{\phi}} , \quad \psi_1(1)=0, \label{eq:PorousFisherStefanSmallc2}
\end{align}
with exact solutions
\begin{align}
\psi_0(\phi)&= \pm \sqrt{\dfrac{2-8\phi^{3/2}+6\phi^2}{3}} ,\label{eq:PorousFisherStefanSmallcsolution1} \\
\psi_1(\phi)&= -\dfrac{1}{\psi_0(\phi)}\int_{\phi}^{1} \sqrt{\dfrac{2-8s^{3/2}+6s^2 }{3s}} \, \textrm{d}s, \label{eq:PorousFisherStefanSmallcsolution2}
\end{align}
where the integral in (\ref{eq:PorousFisherStefanSmallcsolution2}) can be evaluated using quadrature.    Results in Figure \ref{F13} compare a range of phase planes and associated travelling wave profile shapes for $c = \pm 0.1$ and $c = \pm 0.5$.  The upper panel in each subfigure shows the $(\phi,\psi)$ phase plane with a numerical trajectory obtained by solving (\ref{eq:PFSPhasePlaneShape}) using DifferentialEquations.jl in Julia. Each trajectory leaves the $(1,0)$ saddle point along the unstable manifold into the fourth quadrant, eventually intersecting the $\psi$ axis at a special point denoted with the green square.  Each phase plane also includes a plot of the two-term perturbation solution $\psi_0(\phi) + c\psi_1(\phi)$ which gives curves that are indistinguishable from the numerically-generated trajectories for $c = \pm 0.1$ whereas there is a small difference between the two trajectories for $c=\pm 0.5$, as expected.  As for the Fisher-Stefan phase plane analysis in Figure \ref{F10}, we use numerical integration to construct the $\phi(z)$ profile from the phase plane trajectories, and these profiles are then plotted as $U(z) = \sqrt{\phi(z)}$ in the lower panel of each subfigure where we compare the shape of the travelling wave profile obtained using a numerical solution and perturbation solutions of (\ref{eq:PFSPhasePlaneShape}).

Evaluating our two-term perturbation solution at $\phi = 0$ allows us to relate $c$ and $\kappa$ through the moving boundary condition which gives
\begin{equation}
\kappa = \sqrt{6}c + \dfrac{\alpha \sqrt{3}}{27}c^2 + \mathcal{O}(c^3),
\end{equation}
where $\alpha = 36\sqrt{2}-6\sqrt{3}+24\ln\left[(\sqrt{3}-1)/(3\sqrt{2}-4)\right] \approx 67.02$.
A notable feature of the phase planes in Figure \ref{F13} is that $\psi$ remains $\mathcal{O}(1)$ as $\phi \to 0^+$.  Noting that $\phi = U^2$ and $\psi = 2U\textrm{d}U/\textrm{d}z$, then $\textrm{d}U/\textrm{d}z = \mathcal{O}(U^{-1})$ as $U \to 0^{+}$ confirming that $\textrm{d}U/\textrm{d}z$ is infinite at $z=0$.  This property is very different to sharp-fronted solutions of the Porous-Fisher model or the Fisher-Stefan model.  In both these previous cases  $\textrm{d}U/\textrm{d}z$ is finite at the leading edge of the population. This is a new feature that the previous models covered in this review do not share.

\begin{figure}[htp]
\centering
     \includegraphics[width=1.0\textwidth]{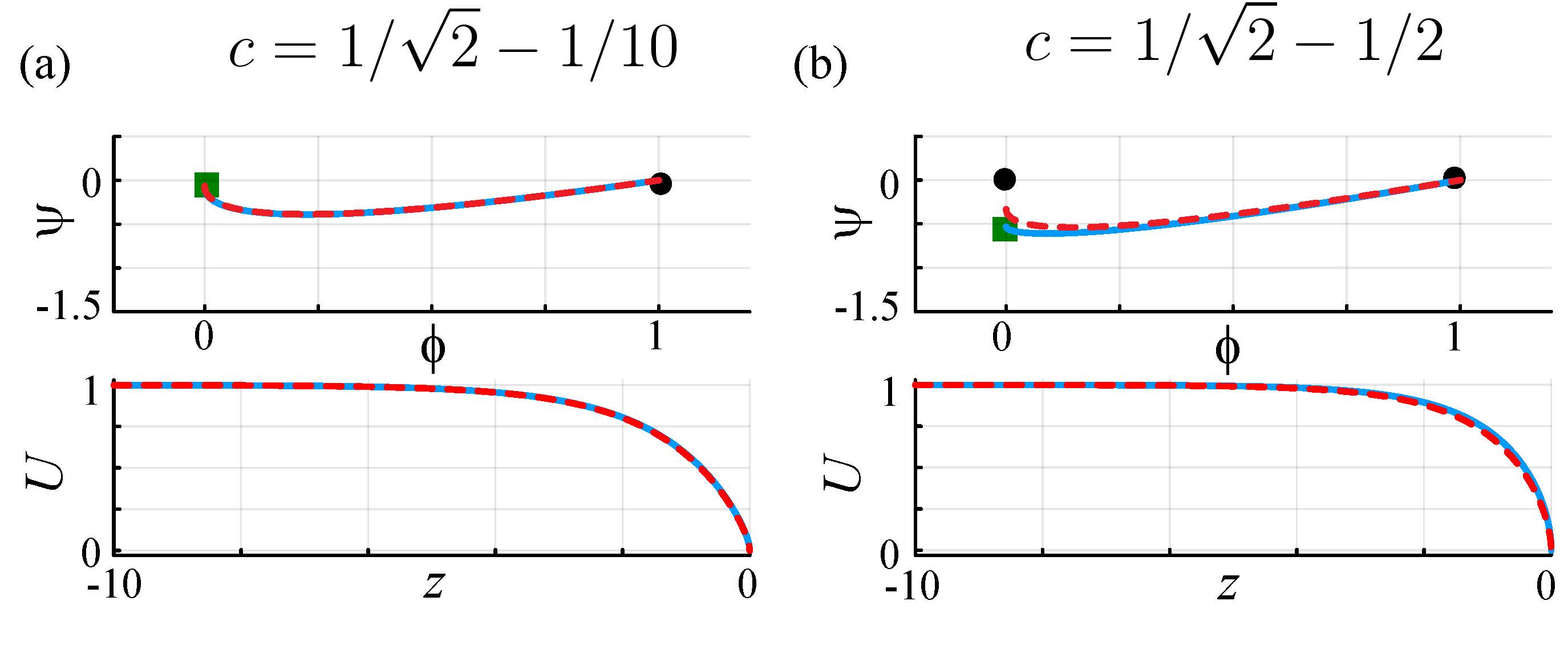}
\renewcommand{\baselinestretch}{1.0}
\caption{Phase planes for (\ref{eq:PhasePlane PFS1})--(\ref{eq:PhasePlane PFS2}) and approximate perturbation solutions for $c = 1/\sqrt{2} - \varepsilon$: (a) $\varepsilon=1/10$; (b) $\varepsilon=1/2$.  The upper panel in both subfigures shows the phase plane where the equilibria are highlighted (black discs) and the special point $(0, \psi^\dagger)$ is also highlighted (green square). The red trajectory between $(1,0)$ and $(0,\psi^\dagger)$ corresponds to the travelling wave solutions and an approximate perturbation solution for $c - 1/\sqrt{2}  \ll 1$ is superimposed.  Results in the lower panel of each subfigure compare numerical estimates of the shape of the travelling wave (solid blue) with solutions obtained from the approximate perturbation solution (dashed red).} \label{F14}
\end{figure}

\newpage
For the special travelling wave speed  $c= 1/\sqrt{2}$, Equation (\ref{eq:PFSPhasePlaneShape}) has the solution that $\textrm{d}U/\textrm{d}z = (U-1)/\sqrt{2}$, meaning that  $\psi =  2U\textrm{d}U/\textrm{d}z \to 0$ as $U \to 0^{+}$.  The moving boundary condition requires that the quantity $-2c/\kappa$ is zero, and since $c > 0$ we must have $\kappa \to \infty$ as $c \to 1/\sqrt{2}^{\, -}$.  The leading-order behaviour of this limit can be explored by writing  $c = 1/\sqrt{2} - \varepsilon$, for $\varepsilon \ll 1$ so that (\ref{eq:PFSPhasePlaneShape}) can be written as
\begin{equation}
\dfrac{\textrm{d}\psi(\phi)}{\textrm{d}\phi} = \dfrac{\varepsilon}{\sqrt{\phi}}-\dfrac{1}{\sqrt{2\phi}}-\dfrac{2\sqrt{\phi}\left(1-\sqrt{\phi}\right)}{\psi(\phi)}, \quad \psi(1) = 0, \quad \lim_{\phi \to 0^+}\psi(\phi) = -\dfrac{2}{\kappa}\left(\dfrac{1}{\sqrt{2}}-\varepsilon  \right).    \label{eq:PFSPhasePlaneShape2}
\end{equation}
Substituting a regular perturbation solution of the form
\begin{equation}
\psi(\phi) = \psi_0(\phi) + \varepsilon \psi_1(\phi) + \mathcal{O}(\varepsilon^2),
\end{equation}
into (\ref{eq:PFSPhasePlaneShape2}) gives,
\begin{align}
\dfrac{\textrm{d}\psi_0(\phi)}{\textrm{d}\phi} &= -\dfrac{1}{\sqrt{2\phi}}-\dfrac{2\sqrt{\phi}\left(1-\sqrt{\phi}\right)}{\psi_0(\phi)}, \quad \psi_0(1)=0, \label{eq:PorousFisherStefanSmallc12}\\
\psi_0(\phi)\dfrac{\textrm{d}\psi_1(\phi)}{\textrm{d}\psi} &=  -\psi_1(\phi)\dfrac{\textrm{d}\psi_0(\phi)}{\textrm{d}\psi} - \dfrac{\psi_1(\phi)}{\sqrt{2\phi}}+\dfrac{\psi_0(\phi)}{\sqrt{\phi}} , \quad \psi_1(1)=0, \label{eq:PorousFisherStefanSmallc2}
\end{align}
with exact solutions
\begin{align}
\psi_0(\phi)&= -\sqrt{2\phi}\left(1-\sqrt{\phi}\right),\label{eq:FisherStefanSmallcsolution12} \\
\psi_1(\phi)&= -\dfrac{2\left(1-\sqrt{\phi}\right)}{3}.\label{eq:FisherStefanSmallcsolution22}
\end{align}
Results in Figure \ref{F14} explore the accuracy of this limit by generating phase planes for $c = 1\/\sqrt{2} - \varepsilon$ for $\varepsilon=0.1, 0.5$ where we see the numerically-generated trajectory obtained by solving (\ref{eq:PFSPhasePlaneShape}) using DifferentialEquations.jl is visually indistinguishable from our two-term perturbation solution for $\varepsilon = 0.1$, whereas we see some small discrepancy between the trajectories for $\varepsilon = 0.5$.  The lower panel in each subfigure compares the numerically-generated phase plane trajectories with the two-term perturbation solution in terms of the $U(z)$ travelling wave profiles.

Evaluating our two term perturbation solution at $\phi=0$ gives
\begin{equation}
\kappa = \dfrac{3\sqrt{2}c}{1-\sqrt{2}c} + \mathcal{O}\left(c - \dfrac{1}{\sqrt{2}}\right)^3
\end{equation}
which shows how $\kappa$ relates to $c$ for $\kappa \gg 1$ and $c \to 1/\sqrt{2}^{\, -}$.

%\disclaimer{Insert disclaimer text here.}

%%%%%%%%%% Insert bibliography here %%%%%%%%%%%%%%

\end{document}